\documentclass[12pt]{article}
\usepackage{amsmath}
\usepackage{amsfonts}
\usepackage{amssymb}
\usepackage{graphicx}
\providecommand{\U}[1]{\protect\rule{.1in}{.1in}}
\providecommand{\U}[1]{\protect\rule{.1in}{.1in}}
\providecommand{\U}[1]{\protect\rule{.1in}{.1in}}
\providecommand{\U}[1]{\protect\rule{.1in}{.1in}}
\providecommand{\U}[1]{\protect\rule{.1in}{.1in}}
\providecommand{\U}[1]{\protect\rule{.1in}{.1in}}
\providecommand{\U}[1]{\protect\rule{.1in}{.1in}}
\providecommand{\U}[1]{\protect\rule{.1in}{.1in}} \textwidth17.1cm \oddsidemargin -0.5cm

\newcommand{\be}{\begin{equation}}
\newcommand{\ee}{\end{equation}}
\newcommand{\ben}{\begin{equation*}}
\newcommand{\een}{\end{equation*}}
\newcommand{\ar}{\begin{array}}
\newcommand{\arn}{\end{array}}

\def\pnot{\mbox{${\not{\hbox{\kern-3.0pt$p$}}}$}}
\def\qnot{\mbox{${\not{\hbox{\kern-2.0pt$q$}}}$}}
\def\enot{\mbox{${\not{\hbox{\kern-2.0pt$e$}}}$}}
\def\knot{\mbox{${\not{\hbox{\kern-2.0pt$k$}}}$}}

\def\fun#1#2{\lower3.6pt\vbox{\baselineskip0pt\lineskip.9pt\ialign
{$\mathsurround=0pt#1\hfil##\hfil$\crcr#2\crcr\sim\crcr}}}

\begin{document}

\begin{titlepage}

\begin{center}
{\bf On the discrepancy of the low-x evolution kernels$^{~\ast}$}
\end{center}
\vskip 0.5cm \centerline{V.S.~Fadin$^{a\,\dag}$,
R.~Fiore$^{b\,\ddag}$, A.V.~Grabovsky$^{a\,\dag\dag}$} \vskip .6cm

\centerline{\sl $^{a}$ Budker Institute of Nuclear Physics, 630090
Novosibirsk, Russia} \centerline{\sl Novosibirsk State University,
630090 Novosibirsk, Russia} \centerline{\sl $^{b}$ Dipartimento di
Fisica, Universit\`a della Calabria,} \centerline{\sl Istituto
Nazionale di Fisica Nucleare, Gruppo collegato di Cosenza,}
\centerline{\sl Arcavacata di Rende, I-87036 Cosenza, Italy}
\vskip 2cm

\begin{abstract}
It is shown  that in the case of the forward scattering the most  part of the
difference between the M\"{o}bius form of  the BFKL kernel and the BK kernel in the next-to-leading order
(NLO)  can be eliminated by the  transformation related to the choice of  the energy scale in the
representation of scattering amplitudes. Change of the nonforward BFKL kernel under this transformation is
derived as well. The functional identity of the forward BFKL kernel in the momentum and M\"{o}bius
representations in the leading order (LO) is exhibited and its NLO validity in $N=4$ supersymmetric
Yang-Mills theory is proved.

\end{abstract}
\vfill \hrule \vskip.3cm \noindent $^{\ast}${\it Work supported in
part by the RFBR grant 07-02-00953, in part by the RFBR--MSTI
grant 06-02-72041, in part by the INTAS
grant 05-1000008-8328 and in part by Ministero Italiano
dell'Istruzione, dell'Universit\`a e della Ricerca.} \vfill $
\begin{array}{ll} ^{\dag}\mbox{{\it e-mail address:}} &
\mbox{FADIN@INP.NSK.SU}\\
^{\ddag}\mbox{{\it e-mail address:}} &
\mbox{FIORE@CS.INFN.IT}\\
^{\dag\dag}\mbox{{\it e-mail address:}} &
\mbox{A.V.GRABOVSKY@INP.NSK.SU}\\
\end{array}
$

\end{titlepage}

\vfill \eject

\section{Introduction}

The Balitsky-Fadin-Kuraev-Lipatov (BFKL) framework  for the
description of semihard processes in QCD was developed originally in the momentum representation. In the
leading order (LO) it was done in Refs.~\cite{BFKL, BL};   the next to leading (NLO) corrections to the
kernel of the BFKL equation were obtained in Refs.~\cite{FL98} - \cite{FF05}. The field of applicability
of the BFKL approach is quite wide. It includes  scattering processes with arbitrary colour exchange.

For scattering of colourless particles  an alternative framework to the  description of the  semihard
processes is  the colour dipole model \cite{dipole}. In contrast to the BFKL approach this model is
formulated in the impact parameter space. The equation describing in the LO evolution of scattering
amplitudes with energy in this model has a simple non-linear generalization, which is called
Balitsky-Kovchegov (BK) equation \cite{Balitsky}.   The NLO corrections to the kernel of the BK equation
in the coordinate space have been  calculated recently in
Refs.~\cite{Kovchegov:2006vj} -\cite{Balitsky:2008zz}.

These two approaches should  give the same predictions in the common area. This requirement is definitely
fulfilled  in the LO. Indeed, for scattering of colourless particles  the LO BFKL kernel can be taken  in
the M\"{o}bius representation, which is invariant with respect to the  conformal transformations of the
transverse coordinates~\cite{Lipatov:1985uk}.  In the impact parameter space the M\"{o}bius representation
of the BFKL kernel  coincides  with the kernel of the colour dipole model~\cite{Fadin:2006ha}
(in Ref.~\cite{Fadin:2006ha} it was called therefore dipole form of the BFKL kernel).   Moreover,  it was
shown \cite{Bartels:2004ef} that the LO BK equation appears as a special case of the nonlinear evolution
equation which sums the fun diagrams for the BFKL Green's functions  in the M\"{o}bius representation.

In the NLO one could  expect coincidence of  the  M\"{o}bius form of the BFKL kernel and the kernel of the
linearized BK equation. However, the situation is not so simple. First of all,  the NLO kernels are not
unambiguously defined. The ambiguity of the NLO kernels is analogous to the ambiguity on the NLO anomalous
dimensions. It  is caused by the possibility to redistribute radiative corrections between the kernels and
the impact factors.  This freedom in the definition of the kernels
allows one to reshape the M\"{o}bius form of the BFKL kernel in order to prove its equivalence to the BK one
\cite{Fadin:2006ha}.

In QCD the  NLO kernels consist of two parts: the quark and the gluon ones. The quark part of the BK
kernel was found in Refs.~\cite{Kovchegov:2006vj} and \cite{Balitsky:2006wa}.  The M\"{o}bius form of the
quark part of
the BFKL kernel  was obtained in Refs.~\cite{Fadin:2006ha} and \cite{Fadin:2007ee}  from the quark
contribution to the
BFKL kernel  calculated in the momentum representation in Ref.~\cite{Fadin:1998jv}.  It was proved
~\cite{Fadin:2006ha,Fadin:2007ee} that with account of the freedom mentioned above this form   is  equivalent
to the quark part of the linearized BK kernel. The M\"{o}bius form of the gluon part   was obtained in
Ref.~\cite{Fadin:2007de} with use of  the  gluon correction  to the BFKL kernel calculated in
Refs.~\cite{FG00} and \cite{FF05} in the momentum representation. The gluon correction to the BK kernel was
found
in Ref.~\cite{Balitsky:2008zz}. It occured that for the linearized BK equation  this correction strongly
differs from the M\"{o}bius form  obtained in  Ref.~\cite{Fadin:2007de}.

In this paper we demonstrate  that for the  case of forward scattering  the most  part of the difference  can
be eliminated by the  transformation related to the choice of  the energy scale in the representation of
scattering amplitudes. We also show how the simplest generalizations  of
this transformation 
change the nonforward kernel.

The second goal of our work is to exhibit the functional identity of the
forward BFKL kernels in the momentum and the M\"{o}bius  representations
and to prove that in the N=4 SUSY Yang-Mills theory this identity remains valid in the NLO. The extension
of the BFKL framework to the supersymmetric theories was started in Ref.~\cite{Kotikov:2000pm},
where the kernel
for the forward scattering was found for the N=4 SUSY in the momentum space with the dimension
$D=4+2\epsilon$ and  in the eigenfunction space.
This analysis has been recently expanded in Ref.~\cite{Fadin:2007xy}, where the
nonforward M\"{o}bius NLO BFKL kernel was obtained for the SUSY Yang-Mills theories with any N. We continue
this line finding the forward kernel in the momentum space for any N, writing it at $D=4$, with all
singularities cancelled, and demonstrating the functional identity of this
kernel to the forward kernel in the M\"{o}bius representation for N=4.

Our paper is organized as follows. The next section introduces our notation
and gives a brief account of the main results of Ref.~\cite{Fadin:2007de}. Section
3 goes through different transforms of the kernel which do not change
observables. Section 4 shows that the difference between the M\"{o}bius form of the BFKL kernel and the
kernel of the linearized BK equation can be partially eliminated by the suitable transform. Section 5
demonstrates the functional identity of the forward BFKL kernels in the momentum and M\"{o}bius coordinate
representations. Section 6 discusses how the generalizations of the
transformation  obtained in 
section 4 change the nonforward kernel. Section 7 summarizes the main
points of our paper. 
Appendices describe the details of the calculations.

\section{General overview}

\label{sec:notation} Our notation is the same as in Refs.~\cite{Fadin:2006ha} and
\cite{Fadin:2007de}. Thus the Reggeon transverse momenta (and the conjugate
coordinates) in initial and final $t$-channel states are $\vec{q}%
_{i}^{\;\prime}\; (\vec{r}_{i}^{\;\prime})$ and $\vec{q}_{i} \; (\vec{r}_{i}),$
$i=1,2$. The state normalization is
\begin{equation}
\langle\vec{q}|\vec{q}^{\;\prime}\rangle=\delta(\vec{q}-\vec{q}^{\;\prime
})\;,\;\;\;\;\;\langle\vec{r}|\vec{r}^{\;\prime}\rangle=\delta(\vec{r}-\vec
{r}^{\;\prime})\;,\quad\quad\langle\vec{r}|\vec{q}\rangle=\frac{e^{i\vec
{q}\,\vec{r}}}{(2\pi)^{1+\epsilon}}.\; \label{normalization}%
\end{equation}
Here $\epsilon=(D-4)/2$; $D-2$ is the transverse space dimension taken
different from $2$ to regularize divergences. We will also
write for brevity $\vec{p}_{ij^{\prime}}=\vec{p}_{i}-\vec{p}_{j}%
^{\;\prime}$.

The $s$-channel discontinuities of scattering amplitudes for the processes
$A+B\rightarrow A^{\prime}+B^{\prime}$ have the form
\begin{equation}
-4i(2\pi)^{D-2}\delta(\vec{q}_{A}-\vec{q}_{B})\mbox{disc}_{s}\mathcal{A}%
_{AB}^{A^{\prime}B^{\prime}}=\langle A^{\prime}\bar{A}|\left(  \frac{s}{s_{0}%
}\right)  ^{\hat{\mathcal{K}}}\frac{1}{\hat{\vec{q}}_{1}^{\;2}\hat{\vec{q}%
}_{2}^{\;2}}|\bar{B}^{\prime}B\rangle\;. \label{discontinuity representation}%
\end{equation}
In this expression $s_{0}$ is an appropriate energy scale, $\hat{\mathcal{K}}%
$\ is the BFKL kernel, $\;\;q_{A}=p_{A^{\prime}A},\;\;q_{B}=p_{BB^{\prime}}$,
and
\begin{equation}
\langle\vec{q}_{1},\vec{q}_{2}|\mathcal{\hat{K}\,}|\vec{q}_{1}^{\,\,\prime
},\vec{q}_{2}^{\,\,\prime}\rangle=\delta(\vec{q}_{1}+\vec{q}_{2}-\vec{q}_{1}^{\;\prime}
-\vec{q}_{2}^{\;\prime})\frac{\mathcal{K}_r(\vec{q}%
_{1},\vec{q}_{1}^{\;\prime};\vec{q})}{\vec{q}%
_{1}^{\,\,2}\vec{q}_{2}^{\,\,2}}+\delta(\vec{q}_{22^{\prime}})\delta\left(
\vec{q}_{11^{\prime}}\right)  \left(  \omega\left(  \vec{q}_{1}^{\,\,2}%
\right)  +\omega(\vec{q}_{2}^{\,\,2})\right)  ,
\label{K_momentum_repr_definition}%
\end{equation}
with $\omega(t)$ being the gluon Regge trajectory, and $\hat{\mathcal{K}}_{r}$
representing real particle production in Reggeon collisions, %
\begin{equation}
\langle\vec{q}_{1},\vec{q}_{2}|\bar{B}^{\prime}B\rangle=4p_{B}^{-}\delta
(\vec{q}_{B}-\vec{q}_{1}-\vec{q}_{2}){\Phi}_{B^{\prime}B}(\vec{q}_{1},\vec
{q}_{2})\;, \label{impact BB}%
\end{equation}%
\begin{equation}
\langle A^{\prime}\bar{A}|\vec{q}_{1},\vec{q}_{2}\rangle=4p_{A}^{+}\delta
(\vec{q}_{A}-\vec{q}_{1}-\vec{q}_{2}){\Phi}_{A^{\prime}A}(\vec{q}_{1},\vec
{q}_{2})\;, \label{impact AA}%
\end{equation}
where $p^{\pm}=(p_{0}\pm p_{z})/\sqrt{2}$; the kernel $\mathcal{K}_r(\vec{q}%
_{1},\vec{q}_{1}^{\;\prime};\vec{q})$ and the impact factors $\Phi$ are
expressed through the Reggeon vertices according to Ref.~\cite{FF98}.

The general
form of the M\"{o}bius (dipole) kernel in the coordinate representation reads
\cite{Fadin:2006ha,Fadin:2007de}:%
\[
\langle\vec{r}_{1}\vec{r}_{2}|\hat{\mathcal{K}}_{M}|\vec{r}_{1}^{\;\prime}%
\vec{r}_{2}^{\;\prime}\rangle=\frac{\alpha_{s}(\mu^{2})N_{c}}{2\pi^{2}}\int
d\vec{\rho}\frac{\vec{r}_{12}{}^{2}}{\vec{r}_{1\rho}^{\,\,2}\vec{r}_{2\rho
}^{\,\,2}}\Biggl[\delta(\vec{r}_{11^{\prime}})\delta(\vec{r}_{2^{\prime}\rho
})+\delta(\vec{r}_{1^{\prime}\rho})\delta(\vec{r}_{22^{\prime}})-\delta
(\vec{r}_{11^{\prime}})\delta({r}_{22^{\prime}})~\Biggr]
+\frac{\alpha_{s}^{2}(\mu^{2})N_{c}^{2}}{4\pi^{3}}
\]%
\begin{equation}
\times\Biggl[\delta(\vec
{r}_{11^{\prime}})\delta(\vec{r}_{22^{\prime}})\int d\vec{\rho}\,g^{0}(\vec
{r}_{1},\vec{r}_{2};\vec\rho)+\delta(\vec{r}_{11^{\prime}})g(\vec{r}_{1},\vec
{r}_{2};\vec{r}_{2}^{\;\prime})+\delta(\vec{r}_{22^{\prime}})g(\vec{r}%
_{2},\vec{r}_{1};\vec{r}_{1}^{\;\prime})+\frac{1}{\pi}g(\vec{r}_{1},\vec
{r}_{2};\vec{r}_{1}^{\;\prime},\vec{r}_{2}^{\;\prime})\Biggr].
\label{K_in_corrdinate_repr}%
\end{equation}
Here $\vec{r}_{i\rho}=\vec{r}_{i}-\vec{\rho}$, and the whole kernel is
symmetric with respect to simultaneous $1\leftrightarrow2$ and $1^{\prime
}\leftrightarrow2^{\prime}$ substitution.

The quark contribution to the functions $g$ was calculated in Refs.~
\cite{Fadin:2006ha} and \cite{Fadin:2007ee} and after a suitable transform was shown to
coincide with the BK result of Ref.~\cite{Balitsky:2006wa}. The gluon contribution to
these functions was found in the BFKL approach in Ref.~\cite{Fadin:2007de} and in
the dipole one in Ref.~\cite{Balitsky:2008zz}. The latter two results are different.

The BFKL framework gives for the gluon contribution%
\begin{equation}
g^{0}(\vec{r}_{1},\vec{r}_{2};\vec{\rho})=-g(\vec{r}_{1},\vec{r}_{2};\vec
{\rho})\ +2\pi\zeta(3)\delta\left(  \vec{\rho}\right)  , \label{g0BFKL}%
\end{equation}%
\[
g(\vec{r}_{1},\vec{r}_{2};\vec{r}_{2}^{\;\prime})\ =\frac{11}{6}\frac{\vec
{r}_{12}^{\;2}}{\vec{r}_{22^{\prime}}^{\;2}\vec{r}_{12^{\prime}}^{\;2}}%
\ln\left(  \frac{\vec{r}_{12}^{\;2}}{r_{\mu}^{2}}\right)  +\frac{11}{6}\left(
\frac{1}{\vec{r}_{22^{\prime}}^{\,\,\,2}}-\frac{1}{\vec{r}_{12^{\prime}%
}^{\,\,\,2}}\right)  \ln\left(  \frac{\vec{r}_{22^{\prime}}^{\,\,\,2}}{\vec
{r}_{12^{\prime}}^{\,\,\,2}}\right)
\]%
\begin{equation}
+\frac{1}{2\vec{r}_{22^{\prime}}^{\;2}}\ln\left(  \frac{\vec{r}_{12^{\prime}%
}^{\;2}}{\vec{r}_{22^{\prime}}^{\;2}}\right)  \ln\left(  \frac{\vec{r}%
_{12}^{\;2}}{\vec{r}_{12^{\prime}}^{\;2}}\right)  -\frac{\vec{r}_{12}^{\;2}%
}{2\,\vec{r}_{22^{\prime}}^{\;2}\vec{r}_{12^{\prime}}^{\;2}}\ln\left(
\frac{\vec{r}_{12}^{\;2}}{\vec{r}_{22^{\prime}}^{\;2}}\right)  \ln\left(
\frac{\vec{r}_{12}^{\;2}}{\vec{r}_{12^{\prime}}^{\;2}}\right)  ,
\label{g3BFKL}%
\end{equation}
where
\begin{equation}
\ln r_{\mu}^{2}=-2C-\ln\frac{\mu^{2}}{4}-\frac{3}{11}\left(  \frac{67}%
{9}-2\zeta(2)\right)
\end{equation}
and $C=-\psi\left(  1\right)  $ is the Euler constant; $\mu$ is a
renormalization scale in the $\overline{MS}$-scheme. At once we will
emphasize that only the integral of $g^{0}(\vec{r}_{1},\vec{r}_{2};\rho)$
contributes to the kernel. Therefore one can write $g^{0}$ in various forms,
e.g. in our previous papers we used the equalities
\begin{equation}
\int d\vec{\rho}\frac{\,\vec{r}_{12}^{\;2}}{\vec{r}_{1\rho}^{\;2}\vec
{r}_{2\rho}^{\;2}}\ln\left(  \frac{\vec{r}_{1\rho}^{\;2}}{\vec{r}_{12}^{\;2}%
}\right)  \ln\left(  \frac{\vec{r}_{2\rho}^{\;2}}{\vec{r}_{12}^{\;2}}\right)
=\int\frac{d\vec{\rho}\,}{\vec{r}_{2\rho}^{\;2}}\ln\left(  \frac{\vec
{r}_{1\rho}^{\;2}}{\vec{r}_{12}^{\;2}}\right)  \ln\left(  \frac{\vec{r}%
_{1\rho}^{\;2}}{\vec{r}_{2\rho}^{\;2}}\right)  =4\pi\zeta(3)
\label{representation for zeta in coordinate space}%
\end{equation}
and
\begin{equation}
\int d\vec{\rho}\left[  \frac{\vec{r}_{12}^{\,\,2}}{\vec{r}_{1\rho}^{\,\,2}%
\vec{r}_{2\rho}^{\,\,\,2}}\ln\left(  \frac{\vec{r}_{1\rho}^{\,\,2}\vec
{r}_{2\rho}^{\,\,\,2}}{\vec{r}_{12}^{\,\,4}}\right)  +\left(  \frac{1}{\vec
{r}_{2\rho}^{\,\,\,2}}-\frac{1}{\vec{r}_{1\rho}^{\,\,\,2}}\right)  \ln\left(
\frac{\vec{r}_{1\rho}^{\,\,2}}{\vec{r}_{2\rho}^{\,\,\,2}}\right)  \right]  =0,
\end{equation}
to reshape it. But anyway $g^{0}(\vec{r}_{1},\vec{r}_{2};\rho)$ does not
coincide with $g(\vec{r}_{1},\vec{r}_{2};\rho)$. For the third function g
we have%
\begin{align}
g(\vec{r}_{1},\vec{r}_{2};\vec{r}_{1}^{\;\prime},\vec{r}_{2}^{\;\prime})  &
=\frac{1}{2\vec{r}_{1^{\prime}2^{\prime}}^{\,\,4}}\left(  \frac{\vec
{r}_{11^{\prime}}^{\;2}\,\vec{r}_{22^{\prime}}^{\;2}-2\vec{r}_{12}^{\;2}%
\,\vec{r}_{1^{\prime}2^{\prime}}^{\;2}}{d}\ln\left(  \frac{\vec{r}%
_{12^{\prime}}^{\;2}\,\vec{r}_{21^{\prime}}^{\;2}}{\vec{r}_{11^{\prime}}%
^{\;2}\vec{r}_{22^{\prime}}^{\;2}}\right)  -1\right)  +\frac{\vec{r}%
\,\,_{12}^{2}\ln\ \left(  \frac{\vec{r}\,\,_{11^{\prime}}^{2}}{\vec
{r}\,\,_{1^{\prime}2^{\prime}}^{2}}\right)  }{2\vec{r}\,\,_{11^{\prime}}%
^{2}\vec{r}\,\,_{12^{\prime}}^{2}\ \vec{r}\,\,_{22^{\prime}}^{2}}\nonumber\\
&  +\frac{\ln\left(  \frac{\vec{r}_{12^{\prime}}^{\;2}\,\vec{r}_{21^{\prime}%
}^{\;2}}{\vec{r}_{11^{\prime}}^{\;2}\vec{r}_{22^{\prime}}^{\;2}}\right)
}{4\vec{r}\,\,_{11^{\prime}}^{2}\ \vec{r}\,\,_{22^{\prime}}^{2}}\left(
\frac{\vec{r}\,\,_{12}^{4}}{d}-\frac{\vec{r}\,\,_{12}^{2}}{\vec{r}%
\,\,_{1^{\prime}2^{\prime}}^{2}}\right)  +\frac{\ln\left(  \frac{\vec
{r}\,\,_{12}^{2}\ \vec{r}\,\,_{1^{\prime}2^{\prime}}^{2}}{\vec{r}%
\,\,_{11^{\prime}}^{2}\vec{r}\,\,_{22^{\prime}}^{2}}\right)  }{2\vec
{r}\,\,_{12^{\prime}}^{2}\vec{r}\,\,_{21^{\prime}}^{2}\ }\left(  \frac{\vec
{r}\,\,_{12}^{2}}{2\vec{r}\,\,_{1^{\prime}2^{\prime}}^{2}}+\frac{1}{2}%
-\frac{\vec{r}\,\,_{22^{\prime}}^{2}}{\vec{r}\,\,_{1^{\prime}2^{\prime}}^{2}%
}\right) \nonumber\\
&  +\frac{\vec{r}\,\,_{21^{\prime}}^{2}\ln\left(  \frac{\vec{r}%
\,\,_{21^{\prime}}^{2}\vec{r}\,\,_{1^{\prime}2^{\prime}}^{2}}{\vec{r}%
\,\,_{12}^{2}\ \vec{r}\,\,_{11^{\prime}}^{2}}\right)  }{2\vec{r}%
\,\,_{11^{\prime}}^{2}\vec{r}\,\,_{22^{\prime}}^{2}\vec{r}\,\,_{1^{\prime
}2^{\prime}}^{2}}+\frac{\ln\ \left(  \frac{\vec{r}\,\,_{12}^{2}}{\vec
{r}\,\,_{1^{\prime}2^{\prime}}^{2}}\right)  }{4\vec{r}\,\,_{11^{\prime}}%
^{2}\vec{r}\,\,_{22^{\prime}}^{2}}+\frac{\ln\left(  \frac{\vec{r}%
\,\,_{22^{\prime}}^{2}}{\vec{r}\,\,_{12}^{2}}\right)  }{2\vec{r}%
\,\,_{11^{\prime}}^{2}\ \vec{r}\,\,_{12^{\prime}}^{2}}+\frac{\vec{r}%
\,\,_{12}^{2}\ln\ \left(  \frac{\vec{r}\,\,_{12}^{2}\vec{r}\,\,_{1^{\prime
}2^{\prime}}^{2}}{\vec{r}\,\,_{12^{\prime}}^{2}\ \vec{r}\,\,_{21^{\prime}}%
^{2}}\right)  }{4\vec{r}\,\,_{11^{\prime}}^{2}\vec{r}\,\,_{22^{\prime}}%
^{2}\vec{r}\,\,_{1^{\prime}2^{\prime}}^{2}}\nonumber\\
&  +\frac{\ln\ \left(  \frac{\vec{r}\,\,_{12}^{2}\vec{r}\,\,_{1^{\prime
}2^{\prime}}^{2}}{\vec{r}\,\,_{12^{\prime}}^{2}\vec{r}\,\,_{22^{\prime}}^{2}%
}\right)  }{2\vec{r}\,\,_{11^{\prime}}^{2}\ \vec{r}\,\,_{1^{\prime}2^{\prime}%
}^{2}}+\frac{\ln\left(  \frac{\vec{r}\,\,_{12}^{2}\vec{r}\,\,_{11^{\prime}%
}^{2}}{\vec{r}\,\,_{22^{\prime}}^{2}\ \vec{r}\,\,_{1^{\prime}2^{\prime}}^{2}%
}\right)  }{2\vec{r}\,\,_{12^{\prime}}^{2}\vec{r}\,\,_{1^{\prime}2^{\prime}%
}^{2}}+(1\leftrightarrow2,1^{\prime}\leftrightarrow2^{\prime}),
\label{gBFKL12}
\end{align}
where
\begin{equation}
d=\vec{r}_{12^{\prime}}^{\;2}\vec{r}_{21^{\prime}}^{\;2}-\vec{r}_{11^{\prime}%
}^{\;2}\vec{r}_{22^{\prime}}^{\,\,2}. \label{gBFKL}%
\end{equation}
The functions $g(\vec{r}_{1},\vec{r}_{2};\vec{r}_{2}^{\;\prime})$ and
$g(\vec{r}_{1},\vec{r}_{2};\vec{r}_{1}^{\;\prime},\vec{r}_{2}^{\;\prime}%
)$\ vanish at $\vec{r}_{1}=\vec{r}_{2}.$ In Ref.~\cite{Fadin:2007de} the latter
function was presented in a form where this property was obvious. For integration,
however, it is more convenient to use the expression (\ref{gBFKL12}) .

At the same time the gluon part of the kernel of the linearized BK equation
gives \cite{Balitsky:2008zz} :%
\begin{equation}
g_{BC}^{0}(\vec{r}_{1},\vec{r}_{2},\vec{\rho})=-g_{BC}(\vec{r}_{1},\vec{r}%
_{2},\vec{\rho})~,
\end{equation}%
\[
g_{BC}(\vec{r}_{1},\vec{r}_{2};\vec{r}_{2}^{\;\prime})\ =\frac{11}{6}%
\frac{\vec{r}_{12}^{\;2}}{\vec{r}_{22^{\prime}}^{\;2}\vec{r}_{12^{\prime}%
}^{\;2}}\ln\left(  \frac{\vec{r}_{12}^{\;2}}{r_{\mu_{B}}^{2}}\right)
+\frac{11}{6}\left(  \frac{1}{\vec{r}_{22^{\prime}}^{\,\,\,2}}-\frac{1}%
{\vec{r}_{12^{\prime}}^{\,\,\,2}}\right)  \ln\left(  \frac{\vec{r}%
_{22^{\prime}}^{\,\,\,2}}{\vec{r}_{12^{\prime}}^{\,\,\,2}}\right)
\]%
\begin{equation}
-\frac{\vec{r}_{12}^{\;2}}{\,\vec{r}_{22^{\prime}}^{\;2}\vec{r}_{12^{\prime}%
}^{\;2}}\ln\left(  \frac{\vec{r}_{12}^{\;2}}{\vec{r}_{22^{\prime}}^{\;2}%
}\right)  \ln\left(  \frac{\vec{r}_{12}^{\;2}}{\vec{r}_{12^{\prime}}^{\;2}%
}\right)  ,
\end{equation}
where
\begin{equation}
\ln r_{\mu_{BC}}^{2}=-\ln\mu^{2}-\frac{3}{11}\left(  \frac{67}{9}%
-2\zeta(2)\right)  .
\end{equation}%
\[
g_{BC}(\vec{r}_{1},\vec{r}_{2};\vec{r}_{1}^{\;\prime},\vec{r}_{2}^{\;\prime
})=\ln\left(  \frac{\vec{r}_{12^{\prime}}^{\,\,2}\vec{r}_{21^{\prime}}%
^{\,\,2}}{\vec{r}_{11^{\prime}}^{\,\,2}\vec{r}_{22^{\prime}}^{\,\,2}}\right)
\left[  \frac{\vec{r}_{11^{\prime}}^{\,\,2}\vec{r}_{22^{\prime}}^{\,\,2}%
+\vec{r}_{12^{\prime}}^{\;2}\vec{r}_{21^{\prime}}^{\;2}-4\vec{r}_{12}%
^{\,\,2}\vec{r}_{1^{\prime}2^{\prime}}^{\,\,2}}{2d\,\vec{r}_{1^{\prime
}2^{\prime}}^{\,\,4}}\right.
\]%
\begin{equation}
\left.  +\frac{1}{4\vec{r}_{11^{\prime}}^{\,\,2}\vec{r}_{22^{\prime}}^{\,\,2}%
}\left(  \frac{\vec{r}_{12}^{\,\,4}}{d}-\frac{\vec{r}_{12}^{\,\,2}}{\vec
{r}_{1^{\prime}2^{\prime}}^{\,\,2}}\right)  +\frac{1}{4\vec{r}_{12^{\prime}%
}^{\,\,2}\vec{r}_{21^{\prime}}^{\,\,2}}\left(  \frac{\vec{r}_{12}^{\,\,4}}%
{d}+\frac{\vec{r}_{12}^{\,\,2}}{\vec{r}_{1^{\prime}2^{\prime}}^{\,\,2}%
}\right)  \right]  -\frac{1}{\vec{r}_{1^{\prime}2^{\prime}}^{\,\,4}}.
\label{gBK}%
\end{equation}

\section{Ambiguity in the definition of the kernel}

To begin with, we are going to discuss the transformations of the kernel which
do not affect observables.

First of all, $\mbox{disc}_{s}\mathcal{A}_{AB}^{A^{\prime}B^{\prime}}$
in Eq.~(\ref{discontinuity representation}) remains intact if one changes both
the kernel and the impact factors via an arbitrary nonsingular operator
$\hat{\mathcal{O}}$:
\begin{equation}
\hat{\mathcal{K}}\rightarrow\hat{\mathcal{O}}^{-1}\hat{\mathcal{K}}%
\hat{\mathcal{O}}~,\;\;\langle A^{\prime}\bar{A}|\rightarrow\langle A^{\prime
}\bar{A}|\hat{\mathcal{O}}~,\;\;\frac{1}{\hat{\vec{q}}_{1}^{\;2}\hat{\vec{q}%
}_{2}^{\;2}}|\bar{B}^{\prime}B\rangle\rightarrow{\hat{\mathcal{O}}^{-1}}%
\frac{1}{\hat{\vec{q}}_{1}^{\;2}\hat{\vec{q}}_{2}^{\;2}}|\bar{B}^{\prime
}B\rangle. \label{kernel transformation}%
\end{equation}
If the LO kernel is  fixed by the requirement that its M\"{o}bius form  coincides with the
kernel of the dipole model, one can shift the NLO contribution
using $\hat{\mathcal{O}}=1-\hat{O}$, with $\hat{O}\sim\alpha_{s}$, and get
\begin{equation}
\hat{\mathcal{K}}\rightarrow\hat{\mathcal{K}}-[\hat{\mathcal{K}}^{(B)},\hat
{O}],  \label{kernel_transforms}%
\end{equation}
where the superscript $(B)$ means the LO kernel. Note  that the M\"{o}bius form calculated in
Refs. \cite{Fadin:2006ha} and \cite{ Fadin:2007de} and presented in the previous section corresponds to the
kernel obtained  by the transformation
\begin{equation}
\hat{\mathcal{K}}\rightarrow\hat{\mathcal{K}}+\frac{\alpha_s}{8\pi}\beta_0
[\hat{\mathcal{K}}^{(B)}, \ln\left(\hat{\vec{q}}_{1}^{\;2}\hat{\vec{q}%
}_{2}^{\;2}\right)
]~ ,  \label{transform with beta}%
\end{equation}
where $\beta_0$ is the first coefficient of the beta-function,
from the kernel defined in Eq.~(\ref{K_momentum_repr_definition}).
This transformation simplifies the M\"{o}bius form and allows to alter the quark part of this form  so
that it agrees with the result of Ref.~\cite{Balitsky:2006wa}.

Secondly, there is a freedom in the energy scale $s_{0}.$  At first sight, it can lead to an additional
ambiguity of the NLO kernel. However, it is not so.  Indeed,  it was shown \cite{Fadin:1998sg}  that any
change of the energy scale can be compensated by the corresponding redefinition of the impact factors. An
experienced reader  can wonder  remembering that in Ref.~\cite{FL98}  the scale transformation was associated
with the change of the kernel. The  matter is that in Ref.~\cite{FL98} one of the impact factors was fixed.
Actually, instead of transforming  both impact factors one can compensate any  change of the scale by
transformation of one of the impact factors and the kernel.  Evidently, in this  case  the transformation of
the kernel has the  form (\ref{kernel_transforms})  with some  specific form of the operator $\hat{O}$. Let
us discuss this question  more in detail.

We begin with the case when  $s_0$ depends only on properties of  scattering particles.  Just this case was
supposed in the definition of the kernel and the impact factors in Ref.~\cite{FF98}. Note that $s_0$ can be
taken as a free parameter.   This freedom  can be used for optimization of perturbative
results~\cite{Ivanov:2005gn}.  A natural choice  is $s_0=Q_AQ_B$, where $Q_A$ and $Q_B$ are typical
virtualities  for the impact  factors ${\Phi}_{A^{\prime}A}$ and ${\Phi}_{B^{\prime}B}$ correspondingly.
Let us consider the transition from such scale to the scale depending on the Reggeon momenta $\vec q_{Ai}$
and $\vec q_{Bi}, \;\;i=1,2$, in the impact factors ${\Phi}_{A^{\prime}A}$ and ${\Phi}_{B^{\prime}B}$
respectively:
\begin{equation}
s_0\rightarrow f_Af_B,  \;\; f_A\equiv f_A(\vec q_{Ai}), \;\; f_B\equiv f_B(\vec q_{Bi}).
\end{equation}
Remind that with the NLO accuracy for any $s$-independent value $c$ one has
\begin{equation}
c^{\,\hat{\mathcal{K}}}=1+\hat{\mathcal{K}}^{(B)} \ln c~.
\end{equation}
Therefore one can write
\begin{equation}
\langle \vec q_{A1},\vec q_{A2}|\left(  \frac{s}{s_0}\right)^{\hat{\mathcal{K}}}
|\vec q_{B1},\vec q_{B2}\rangle =
\langle \vec q_{A1},\vec q_{A2}|\hat F_A\left(  \frac{s}{f_Af_B}\right)^{\hat{\mathcal{K}}}\hat F_B
|\vec q_{B1},\vec q_{B2}\rangle,
\label{s0 to fAfB}%
\end{equation}
where
\[
\hat F_A=\left(1+\ln\left(\frac{\hat f_A}{s_0^{\alpha}}\right)\hat{\mathcal{K}}^{(B)}\right), \;\;
\hat F_B=\left(1+\hat{\mathcal{K}}^{(B)}\ln\left(\frac{\hat f_A}{s_0^{\beta}}\right)\right),\;\;
\]
\begin{equation}
\alpha+\beta=1, \;\;\hat f_A\equiv f_A(\hat{\vec q}_{i}), \;\; \hat f_B\equiv f_B(\hat{\vec q}_{i}).
\end{equation}
It means that the discontinuity (\ref{discontinuity representation}) remains
unchanged if the change of the scale (\ref{s0 to fAfB}) is accompanied by the change
of the impact factors
\begin{equation}
\langle A^{\prime}\bar{A}|\rightarrow \langle A^{\prime}\bar{A}|\hat F_A, \;\;
\frac{1}{\hat{\vec{q}}_{1}^{\;2}\hat{\vec{q}}_{2}^{\;2}}|\bar{B}^{\prime
}B\rangle\rightarrow \hat F_B\frac{1}{\hat{\vec{q}}_{1}^{\;2}\hat{\vec{q}}_{2}^{\;2}}|\bar{B}^{\prime
}B\rangle.\label{IF transform}
\end{equation}%
It is possible to leave one of the impact factors (let us take for definiteness
$\langle A^{\prime}\bar{A}|$) invariable  changing the kernel.
Indeed,
\begin{equation}
\hat F_A \left(  \frac{s}{
{f}_{A}{f}_B}\right)^{\hat{\mathcal{K}}}\hat F_A^{\,-1}= \left(  \frac{s}{
{f}_{A}{f}_B}\right)^{\hat{\mathcal{K}}'}, \;\; \hat{\mathcal{K}}' =
\hat F_A \hat{\mathcal{K}}\hat F_A^{\,-1}~.
\end{equation}
Therefore instead of the change (\ref{IF transform}) one can take
\begin{equation}
\hat{\mathcal{K}}\rightarrow \hat{\mathcal{K}}', \;\;
\frac{1}{\hat{\vec{q}}_{1}^{\;2}\hat{\vec{q}}_{2}^{\;2}}|\bar{B}^{\prime
}B\rangle\rightarrow \hat F_A\hat F_B\frac{1}{\hat{\vec{q}}_{1}^{\;2}\hat{\vec{q}}_{2}^{\;2}}|\bar{B}^{\prime
}B\rangle, \label{K and IF transform}
\end{equation}%
where, with the NLO accuracy,
\begin{equation}
\hat{\mathcal{K}}'=\hat{\mathcal{K}}-\left[
\hat{\mathcal{K}}^{(B)},\ln\hat{f}_{A}\hat{\mathcal{K}}^{(B)}\right]~.\label{K tran}
\end{equation}
We see that the change of the  energy scale   can be associated with the transformation of   the
kernel ~(\ref{kernel_transforms})  with the specific $\hat{O}$.

At last, the M\"{o}bius kernel (\ref{K_in_corrdinate_repr}) is defined with an
accuracy to any function independent of $\vec{r}_{1}$ or of $\vec{r}_{2}$
such that after their addition the kernel remains zero at $\vec{r}_{1}=\vec
{r}_{2}$ $\cite{Fadin:2006ha,Fadin:2007de}. $ Therefore, one can add to the
kernel only the functions which are antisymmetric with respect to the substitution $\vec{r}%
_{1}^{\,\,\prime}\leftrightarrow\vec{r}_{2}^{\,\,\prime}$. These
functions do not change the symmetric part of the kernel. But this part alone
plays a role in the observables. As a result, the third term in the expression
(\ref{gBFKL12}) in the BFKL kernel,
\begin{equation}
\frac{\ln\left(  \frac{\vec{r}_{12^{\prime}}^{\;2}\,\vec{r}_{21^{\prime}%
}^{\;2}}{\vec{r}_{11^{\prime}}^{\;2}\vec{r}_{22^{\prime}}^{\;2}}\right)
}{4\vec{r}\,\,_{11^{\prime}}^{2}\ \vec{r}\,\,_{22^{\prime}}^{2}}\left(
\frac{\vec{r}\,\,_{12}^{4}}{d}-\frac{\vec{r}\,\,_{12}^{2}}{\vec{r}%
\,\,_{1^{\prime}2^{\prime}}^{2}}\right)  +(1\leftrightarrow2,1^{\prime
}\leftrightarrow2^{\prime})
\end{equation}
and the term
\begin{equation}
\ln\left(  \frac{\vec{r}_{12^{\prime}}^{\,\,2}\vec{r}_{21^{\prime}}^{\,\,2}%
}{\vec{r}_{11^{\prime}}^{\,\,2}\vec{r}_{22^{\prime}}^{\,\,2}}\right)  \left[
\frac{1}{4\vec{r}_{11^{\prime}}^{\,\,2}\vec{r}_{22^{\prime}}^{\,\,2}}\left(
\frac{\vec{r}_{12}^{\,\,4}}{d}-\frac{\vec{r}_{12}^{\,\,2}}{\vec{r}_{1^{\prime
}2^{\prime}}^{\,\,2}}\right)  +\frac{1}{4\vec{r}_{12^{\prime}}^{\,\,2}\vec
{r}_{21^{\prime}}^{\,\,2}}\left(  \frac{\vec{r}_{12}^{\,\,4}}{d}+\frac{\vec
{r}_{12}^{\,\,2}}{\vec{r}_{1^{\prime}2^{\prime}}^{\,\,2}}\right)  \right]
\end{equation}
in the BK kernel (\ref{gBK}) give the same contribution to the amplitudes.

  The ambiguities  of the NLO kernels give a hope that  the results of
Refs. \cite{Fadin:2007de} and \cite{Balitsky:2008zz}  can be matched.

\section{The kernel for forward scattering in gluodynamics}

For a start let us find the gluon contribution to the forward M\"{o}bius BFKL
kernel and compare it to the BK result obtained in Ref.~\cite{Balitsky:2008zz}. We
define the matrix element of the forward kernel in the momentum representation as
\begin{equation}
\langle\vec{q}|\hat{\mathcal{K}}|\vec
{q}^{\,\,\prime}\rangle=\int\langle\vec{q}, \vec{l}|\mathcal{\hat
{K}\,}|\vec{q}^{\,\,\prime},-\vec{q}^{\,\,\prime}\rangle d \vec{l}.
\end{equation}
Next, using  the  state normalization (\ref{normalization}) and the denotation
$\vec{r}_{1}=\vec r +\vec{r}_{2}, \;\;\vec{r}_{1}^{\,\,\prime}=\vec r^{\,\,\prime} +
\vec{r}_{2}^{\,\,\prime}$,
 we get at physical value $D=4$%
\begin{equation}
\langle\vec{q}|\hat{\mathcal{K}}|\vec
{q}^{\,\,\prime}\rangle=\int \frac{d\vec{r}}{2\pi}\frac{d\vec r^{\,\,\prime}}{2\pi}
e^{-i \vec{q}\vec{r}+i\vec{q}^{\,\,\prime}\vec r^{\,\,\prime}}
\langle\vec{r}|\hat{\mathcal{K}}|\vec
{r}^{\,\,\prime}\rangle~, \label{forward_kernel_momentum}
\end{equation}
where
\begin{equation}
\langle\vec{r}|\hat{\mathcal{K}}|\vec
{r}^{\,\,\prime}\rangle=\int d\vec{r}_{2}^{\,\,\prime}\langle\vec{r}, \vec{0}|
\hat{\mathcal{K}}|\vec{r}^{\;\prime}+\vec{r}_{2}^{\;\prime}, \vec{r}_2^{\;\prime}\rangle
=\int\langle\vec{r}_{1}\vec
{r}_{2}|\hat{\mathcal{K}}|\vec{r}_{1}^{\;\prime}\vec{r}_{2}^{\;\prime}%
\rangle\delta(\vec{r}_{1^{\prime}2^{\prime}}-\vec{r}^{\;\prime})d\vec{r}_{1}^{\;\prime
}d\vec{r}_{2}^{\;\prime}~.\label{forward_kernel_coord}
\end{equation}
The last equality follows from the space uniformity.

Thus, the  M\"{o}bius form of the BFKL kernel for the forward scattering can be
obtained   from  Eq.~(\ref{forward_kernel_coord}) using the results of
Refs.~\cite{Fadin:2006ha}, \cite{Fadin:2007ee} and \cite{Fadin:2007de} for  $\langle\vec{r}_{1}\vec
{r}_{2}|\hat{\mathcal{K}}_M|\vec{r}_{1}^{\;\prime}\vec{r}_{2}^{\;\prime}\rangle $.
In the case of pure gluodynamics it gives

\[
\langle\vec{r}|\hat{\mathcal{K}}_M|\vec{r^{\;\prime}}\rangle
=\frac{\alpha_{s}(\frac{4e^{-2C}}{\vec{r}^{\:2}})N_{c}}{2\pi^2}\int
\frac{d\vec{\rho}\;\vec{r}^{\;2}}{(\vec{r}
-\vec{\rho})^{2}\vec{\rho}^{\:2}}\left\{
\left(2\delta(\vec{\rho}-\vec{r}^{\;\prime}) -\delta(\vec{r}-\vec{r^{\;\prime}})\right)\left[1+
\frac{\alpha_sN_c}{4\pi}
\left(\frac{67}{9}-2\zeta(2)\right.\right.\right.
\]
\[
\left.\left.\left.
+\frac{11}{3}\frac{\vec{\rho}^{\:2}-(\vec{r}-\vec{\rho})^{2}}{\vec{r}^{\;2}}
\ln\left(  \frac{(\vec{r}-\vec{\rho})^{2}}{\vec{\rho}^{\; 2}}\right)\right)
\right] + \frac{\alpha_sN_c}{4\pi}{3}\delta(\vec{r}-\vec{r^{\;\prime}})\ln\left(  \frac{(\vec{r}-
\vec{\rho})^2}{\vec{r}^{\;2}}\right)  
\ln\left(  \frac{\vec{\rho}^{\:2}}{\vec{r}^{\;2}}\right) \right\}
\]
\begin{equation}
+\frac{\alpha^2_{s}N^2_{c}}{4\pi^3}
\frac{\vec{r}^{\;2}}{\vec{r}^{\;\prime 2}}\left(f_1(\vec{r},\vec{r}^{\;\prime})
+f_2(\vec{r},\vec{r}^{\;\prime}) -\frac{1}{(\vec{r}-\vec{r}^{\;\prime})^{2}}%
\ln^{2}\left(  \frac{\vec{r}^{\;2}}{\vec{r}^{\;\prime 2}}\right)\right) ~. \label{KM}
\end{equation}
Here
\begin{equation}
\alpha_{s}(\frac{4e^{-2C}}{\vec{r}^{\:2}}) \simeq\alpha_{s}(\mu^{2})\left(  1-\frac{\alpha_{s}(\mu^{2}%
)}{4\pi}\frac{11}{3}N_{c}\ln\left(  \frac{4e^{-2C}}{\vec{r}^{\:2}\mu^{2}}\right)
\right)~,
\end{equation}
$\mu$ is the renormalization scale in the $\overline{MS}$-scheme,
\[
f_{1}\left( \vec x,\vec y\right)
=\frac{\left( \vec x^{\;2} -\vec y^{\;2}\right)  }{\left(  \vec x-\vec y\right)  ^{2}\left(
\vec x+\vec y\right)^{2}}\left[  \ln\left(  \frac{\vec x^{\;2}}{\vec y^{\;2}}\right)  \ln\left(
\frac{\vec x^{\;2}\vec y^{\;2}\left(  \vec x-\vec y\right)^{4}}
{\left(  \vec x^{\;2}+\vec y^{\;2}\right)  ^{4}%
}\right)  +2\operatorname{Li}_{2}\left(  -\frac{\vec y^{\;2}}{\vec x^{\;2}}\right)
-2\operatorname{Li}_{2}\left(  -\frac{\vec x^{\;2}}{\vec y^{\;2}}\right)  \right]
\]%
\begin{equation}
-\left(  1-\frac{\left(  \vec x^{\;2}-\vec y^{\;2}\right)^{2}}{\left(  \vec x-\vec y\right)
^{2}\left(  \vec x+\vec y\right)  ^{2}}\right)  \left[  \int_{0}^{1}-\int_{1}^{\infty
}\right]  \frac{du}{\left(  \vec x-\vec yu\right)  ^{2}}
\ln\left(  \frac{u^{2}\vec y^{\;2}%
}{\vec x^{\;2}}\right)  ~, \label{f1}%
\end{equation}%
\[
f_{2}\left(  \vec x,\vec y\right)
=\frac{1}{8\vec x^{\;2}\vec y^{\;2}}\left\{  \left(  \vec x\,\vec y\right)^{2}
\left(  1-\frac{3}%
{2}\left(  \frac{\vec y^{\;2}}{\vec x^{\;2}}+\frac{\vec x^{\;2}}{\vec y^{\;2}}\right)  \right)  +\left(
\vec x^{\;2}+\vec y^{\;2}\right)  ^{2}-32\vec x^{\;2}\vec y^{\;2}\right\}
\int_{0}^{\infty}dt\frac
{\ln\left\vert \frac{1+t}{1-t}\right\vert }{\vec y^{\;2}+t^{2}\vec x^{\;2}}%
\]%
\begin{equation}
+\frac{3\left(  \vec x\,\vec y\right)  ^{2}-2\vec x^{\;2}\vec y^{\;2}}{16\vec x^{\;2}\vec y^{\;2}}
\left(  \ln\frac{\vec x^{\;2}}{\vec y^{\;2}}\left(  \frac{1}{\vec y^{\;2}}-\frac{1}{\vec x^{\;2}}\right)
 +\frac{2}{\vec x^{\;2}}+\frac{2}{\vec y^{\;2}}\right)  ~. \label{f2}%
\end{equation}
The details of the derivation are given in the appendix A.

The result of Ref.~\cite{Balitsky:2008zz} for the forward case is
\[
\langle\vec{r}|\hat{\mathcal{K}}_{BC}|\vec{r^{\;\prime}}\rangle
=\frac{\alpha_{s}(\frac{1}{\vec{r}^{\:2}})N_{c}}{2\pi^2}\int
\frac{d\vec{\rho}\;\vec{r}^{\;2}}{(\vec{r}
-\vec{\rho})^{2}\vec{\rho}^{\:2}}
\left(2\delta(\vec{\rho}-\vec{r}^{\;\prime}) -\delta(\vec{r}-\vec{r^{\;\prime}})\right)\left[1+
\frac{\alpha_sN_c}{4\pi}
\left(\frac{67}{9}-2\zeta(2)\right.\right.
\]
\[
\left.\left.
+\frac{11}{3}\frac{\vec{\rho}^{\:2}-(\vec{r}
-\vec{\rho})^{2}}{\vec{r}^{\;2}}\ln\left(  \frac{(\vec{r}-\vec{\rho})^{2}}{\vec{\rho}^{\; 2}}\right)
-2\ln\left(\frac{(\vec{r}-\vec{\rho})^2}{\vec{r}^{\;2}}\right)  \ln\left(\frac{\vec{\rho}^{\:2}}
{\vec{r}^{\;2}}\right) \right)\right]
\]
\begin{equation}
+\frac{\alpha^2_{s}N^2_{c}}{4\pi^3}
\frac{\vec{r}^{\;2}}{\vec{r}^{\;\prime 2}}\left(f_1(\vec{r},\vec{r}^{\;\prime})
+f_2(\vec{r},\vec{r}^{\;\prime}) \right) ~.
\end{equation}
It was derived using the relation (\ref{forward_kernel_coord}). Note that in the derivation the term
\[
\frac{\alpha^2_{s}N^2_{c}}{4\pi^3}
\frac{\vec{r}^{\;2}}{\vec{r}^{\;\prime 2}}\left(f_1(\vec{r},\vec{r}^{\;\prime})
-f_1(\vec{r},-\vec{r}^{\;\prime})\right)
\]
was omitted. As we  discussed at the end of the previous section, it can be done since this term  vanishes
at $\vec{r}^{\,\,\prime}\leftrightarrow-\vec{r}^{\,\,\prime}$ (i.e. $\vec
{r}_{1}^{\,\,\prime}\leftrightarrow\vec{r}_{2}^{\,\,\prime})$.

Thus, for the difference of the forward kernels  one has
\[
\langle\vec{r}|\hat{\mathcal{K}}_M-\hat{\mathcal{K}}_{BC}|\vec{r^{\;\prime}}\rangle
=\frac{\alpha^2_{s}N^2_{c}}{4\pi^3}\left[  \frac{\,\vec{r}^{\;2}}
{(\vec{r}-\vec{r}^{\;\prime})^{2}\vec{r}^{\;\prime 2}}\ln\left( \frac{\vec{r}^{\;2}}{\vec{r}
^{\;\prime 2}}\right)\ln\left( \frac{\vec{r}^{\;2}\vec{r}
^{\;\prime 2}}
{(\vec{r}-\vec{r}^{\;\prime})^ 4}\right)+\delta(\vec{r}-\vec{r^{\;\prime}})2\pi\zeta(3)\right]
\]
\begin{equation}
+\frac{\alpha_{s}N_{c}}{4\pi}\frac{11}{3}\left(C- \ln2\right)\langle\vec{r}|
\hat{\mathcal{K}}_M^{(B)}|\vec{r^{\;\prime}}\rangle~, \label{K M - K BC}
\end{equation}
where
\begin{equation}
\langle\vec{r}|\hat{\mathcal{K}}_M^{(B)}|\vec{r^{\;\prime}}\rangle
=\frac{\alpha_{s}N_{c}}{2\pi^2}\int
\frac{d\vec{\rho}\;\vec{r}^{\;2}}{(\vec{r}
-\vec{\rho})^{2}\vec{\rho}^{\:2}}
\left(2\delta(\vec{\rho}-\vec{r}^{\;\prime}) -\delta(\vec{r}-\vec{r^{\;\prime}})\right)~.
\label{K M B}
\end{equation}
The term proportional to $11/3$ is related to renormalization (remind that in pure gluodynamics
$\beta_0=11N_c/3$). In our opinion, this term arose because  the renormalization scheme used in
Ref.~\cite{Balitsky:2008zz} is not equivalent to
conventional $\overline{MS}$-scheme.

It occurs that the term with logarithms can be eliminated by the transformation
\begin{equation}
\hat{\mathcal{K}}\rightarrow \hat{\mathcal{K}}+\frac12\left[
\hat{\mathcal{K}}^{(B)},\,\ln(\hat{\vec{q}}^{\,\,2})\,\hat{\mathcal{K}}^{(B)}\right]
~\label{elimination}
\end{equation}
applied to the forward BFKL kernel. Indeed, the direct calculation given in
the appendix B shows that
\begin{equation}
\langle\vec{r}|\left[  \mathcal{\hat{K}}^{(B)},\,\ln(\hat{\vec{q}}^{\,\,2})\,
\mathcal{\hat{K}}^{(B)}\right]_{M}|\vec{r}^{\;'}\rangle=-\frac{\alpha_{s}^{2}
N_{c}^{2}}{2\pi^{3}} \frac{\,\vec{r}^{\;2}}
{(\vec{r}-\vec{r}^{\;\prime})^{2}\vec{r}^{\;\prime 2}}\ln\left( \frac{\vec{r}^{\;2}}{\vec{r}
^{\;\prime 2}}\right)\ln\left( \frac{\vec{r}^{\;2}\vec{r}
^{\;\prime 2}}
{(\vec{r}-\vec{r}^{\;\prime})^ 4}\right). \label{[K_K_lnq1q2]}%
\end{equation}
Here the subscript $M$ means the  M\"{o}bius representation (i.e. vanishing of the matrix element at
$\vec{r}=0$). Comparing with Eq.~(\ref{K tran}) we see that the
transformation (\ref{elimination}) corresponds to
the change of the energy scale at fixed value of one of the impact factors.
Actually,  the transformation (\ref{elimination})  is of the same type as in
Ref.~\cite{FL98}:
\begin{equation}
\mathcal{K}(\vec{q},\vec{q}^{\,\,\prime})\rightarrow\mathcal{K}(\vec
{q},\vec{q}^{\,\,\prime})+\frac{1}{2}\int d\,{\vec{p}}%
\,\,\,\mathcal{K}^{(B)}(\vec{q},\vec{p})\,\ln{\frac{\vec{p}{}^{\,2}%
}{\vec{q}^{\,\,2}}}\mathcal{K}^{(B)}(\vec{p},\vec{q}^{\,\,\prime}).
\label{FL98_forward_kernel_transform}%
\end{equation}
One can come to the transformation (\ref{elimination}) from another side.
The  difference $\hat{\mathcal{K}}_M-\hat{\mathcal{K}}_{BC}$ has the same eigenfunctions
\begin{equation}
\langle \vec r^{\;\prime}|n, \gamma\rangle \sim e^{in\phi_{\vec r^{\;\prime}}}
\left(\vec r^{\;\prime 2}\right)^{\gamma}~,
\label{eigenfunctions in r}%
\end{equation}
where $\phi$  is the azimuthal angle, as the LO dipole kernel.
The eigenvalues  of the LO dipole kernel coincide with the eigenvalues of the LO BFKL kernel obtained in
Ref.~\cite{BL}:
\begin{equation}
\omega_B(n, \gamma)=\frac{\alpha_sN_c}{\pi}\chi(n, \gamma)~, \;\;
\chi(n, \gamma)=2\psi(1)-\psi(\gamma+\frac{n}{2})-\psi(1-\gamma+\frac{n}{2})~.
\label{LO eigenvalues}%
\end{equation}
It becomes evident if we write the FBKL kernel for the forward scattering as
\begin{equation}
\langle\vec{q}|\hat{\mathcal{K}}^{(B)}|\vec{q^{\;\prime}}\rangle =
\frac{\alpha_{s}N_{c}}{2\pi^{2}}\left[\frac{2\vec{q}^{\;\prime 2}}
{(\vec{q}-\vec{q}^{\;\prime})^{2}\,\vec{q}^{\;2}}-\delta(\vec{q}
-\vec{q}^{\;\prime})\int \frac{d\vec{l}\;\vec{q}^{\;2}}{(\vec{q}-\vec{l})^{2}
\vec{l}^{\:2}}\right]
\label{LO BFKL_forward_kernel}%
\end{equation}
and compare it with the dipole kernel
\begin{equation}
\langle\vec{r}|\hat{\mathcal{K}}_d|\vec{r^{\;\prime}}\rangle =
\frac{\alpha_{s}N_{c}}{2\pi^{2}}\left[\frac{2\vec{r}^{\; 2}}
{(\vec{r}-\vec{r}^{\;\prime})^{2}\,\vec{r}^{\;\prime2}}-\delta(\vec{r}
-\vec{r}^{\;\prime})\int \frac{d\vec{\rho}\;\vec{r}^{\;2}}{(\vec{r}-\vec{\rho})^{2}
\vec{\rho}^{\:2}}\right]~.
\label{LO _forward_dipole kernel}%
\end{equation}
It is worthwhile to note here that the kernel (\ref{LO BFKL_forward_kernel}) differs from the usually used
symmetric kernel;  the  former is obtained from the latter by the transformation
$\hat{\mathcal{K}}\rightarrow \hat{\vec{q}}^{\;-2}\hat{\mathcal{K}}\hat{\vec{q}}^{\;2}$. For the non-forward
case the corresponding transformation is $\hat{\mathcal{K}}\rightarrow (\hat{\vec{q}}_1^{\;2}
\hat{\vec{q}}_2^{\;2})^{-1/2}\hat{\mathcal{K}}
(\hat{\vec{q}}_1^{\;2}\hat{\vec{q}}_2^{\;2})^{1/2}$. Let us stress that just the
transformed kernel can be written in the M\"{o}bius form $\langle\vec{r}_{1}\vec{r}_{2}
|\hat{\mathcal{K}}^{(B)}_{M}|\vec{r}_{1}^{\;\prime}\vec{r}_{2}^{\;\prime}\rangle$,
which is invariant in regard to the  conformal
transformations of the transverse coordinates \cite{Lipatov:1985uk} and coincides
with the kernel of the colour dipole model $\langle\vec{r}_{1}\vec{r}_{2}|
\hat{\mathcal{K}}_{d}|\vec{r}_{1}^{\;\prime}%
\vec{r}_{2}^{\;\prime}\rangle$ \cite{Fadin:2006ha}. Moreover,
because of this coincidence, in the forward case one can see from Eqs.~(\ref{LO BFKL_forward_kernel}) and
(\ref{LO _forward_dipole kernel}) the functional identity of the LO BFKL kernel in the momentum  and
M\"{o}bius coordinate representations:
$\vec{q}^{\;2}\langle\vec{q}|\hat{\mathcal{K}}^{(B)}|\vec{q^{\;\prime}}\rangle/\vec{q}^{\;\prime 2}$ is
represented by the same function as $\vec{r}^{\;\prime 2}\langle\vec{r}|\hat{\mathcal{K}}^{(B)}_M|
\vec{r^{\;\prime}}\rangle/\vec{r}^{\;2}$.

The eigenvalues $\omega_B(n, \gamma)$ are  associated usually with the eigenfunctions
$e^{in\phi_{\vec q^{\;\prime}}}\left(\vec q^{\;\prime 2}\right)^{\gamma-2}$ in the momentum space, i.e.
$e^{in\phi_{\vec r^{\;\prime}}}
\left(\vec r^{\;\prime 2}\right)^{1-\gamma}$ in the coordinate space, so that the eigenvalues for
Eq.~(\ref{eigenfunctions in r}) should be $\omega_B(n, 1-\gamma)$.   On the other hand, from the functional
identity of Eqs.~(\ref{LO BFKL_forward_kernel}) and (\ref{LO _forward_dipole kernel}) it is clear  that the
eigenvalues must be the same as for $e^{in\phi_{\vec r^{\;\prime}}}
\left(\vec q^{\;\prime 2}\right)^{\gamma-2}$, i.e. $\omega_B(n, \gamma)$. Both requirements are fulfilled
because  $\omega_B(n, \gamma)=\omega_B(n, 1-\gamma)$.

Using the integral (the calculation is discussed in the Appendix C)
\begin{equation}
\int \frac{d\vec{r}^{\;\prime}}{2\pi}e^{in(\phi_{\vec r^{\;\prime}}-\phi_{\vec r})}
\left(\frac{\vec r^{\;\prime 2}}{\vec{r}^{\;2}}\right)^{\gamma} \frac{\,\vec{r}^{\;2}}
{(\vec{r}-\vec{r}^{\;\prime})^{2}\vec{r}^{\;\prime 2}}\ln\left( \frac{\vec{r}^{\;2}}
{\vec{r}^{\;\prime 2}}\right)\ln\left( \frac{\vec{r}^{\;2}\vec{r}
^{\;\prime 2}}
{(\vec{r}-\vec{r}^{\;\prime})^ 4}\right) =2\chi^{\prime}\left(  n,\gamma\right)
\chi\left(  n,\gamma\right)~, \label{difference in omega}
\end{equation}
where $\chi^{\prime}$ means derivative over $\gamma$, from Eq.~(\ref{K M - K BC}) we obtain
\begin{equation}
\omega_M\left(  n,\gamma\right) -\omega_{BC}\left(  n,\gamma\right)
=\frac{\alpha_{s}^{2}(\mu^{2})N_{c}^{2}}{2\pi^{2}}\left[   \chi^{\prime}\left(  n,\gamma\right)
\chi\left(  n,\gamma\right)
+\frac{11}{3}\left(C-  \ln2\right)  \chi(n,\gamma)+\zeta(
3) \right]~,\label{om-om}
\end{equation}
where $\omega_M\left(  n,\gamma\right) -\omega_{BC}\left(  n,\gamma\right)$ is  the eigenvalue of the
difference $\hat{\mathcal{K}}_M-\hat{\mathcal{K}}_{BC}$ corresponding to the eigenfunction
$e^{in\phi_{\vec r^{\;\prime}}}
\left(\vec r^{\;\prime 2}\right)^{\gamma}$. The first term in Eq.~(\ref{om-om}) can be written as
\begin{equation}
\frac12\omega_B^{\prime}\left(  n,\gamma\right) \omega_B\left(  n,\gamma\right)
=-\frac12 [\omega_B, \frac{\partial}{\partial \gamma}\;\omega_B].
\end{equation}
In the space of the eigenfunctions $e^{in\phi_{\vec r^{\;\prime}}}
\left(\vec r^{\;\prime 2}\right)^{\gamma}$  we have $\hat{\mathcal{K}}^{(B)} =
\omega_B\left(  n,\gamma\right) $ and $\ln (\hat{\vec q}^{\,\,2})=-{\partial}/{\partial \gamma}$, so that
  for the forward scattering we obtain
\begin{equation}
\hat{\mathcal{K}}_M-\hat{\mathcal{K}}_{BC}
=\frac12 [\hat{\mathcal{K}}^{(B)}, \ln (\hat{\vec q}^{\,\,2})\hat{\mathcal{K}}^{(B)}] +
\hat{\mathcal{K}}^{(B)}\frac{11}{3}\frac{\alpha_{s}(\mu^{2})N_{c}}{2\pi} \left(C-\ln2\right) +
\frac{\alpha_{s}^{2}(\mu^{2})N_{c}^{2}}{2\pi^{2}} \zeta(3).\label{k-k}
\end{equation}
Evidently, the first term in Eq.~(\ref{k-k}) is eliminated by the transformation (\ref{elimination}).
The second one,  as it was already pointed out, in our opinion  is related to the difference of  the
renormalization scheme used in Ref.~\cite{Balitsky:2008zz} with conventional $\overline{MS}$-scheme and
can be eliminated by change of the scheme.\footnote{We have to note that in fact this term is present in 
the difference between the eigenvalues  of the NLO BFKL kernel and the linearized forward kernel of
Ref.~\cite{Balitsky:2008zz}. In the calculation of this difference presented in Ref.~\cite{Balitsky:2008zz} 
this term is erroneously omitted at  the transition from Eq.~(120) to Eq.~(122).}  Unfortunately, we cannot 
find the  transformation suitable to eliminate the third term.
We have to add that  in the BFKL approach the term with $\zeta(3)$ passed through a great
number of verifications. In particular, this term  is necessary for the fulfillment of the bootstrap relations.
Besides, it is confirmed by the calculation of the three-loop anomalous dimensions in
Refs.~\cite{Vogt:2004mw} and \cite{Kotikov:2004er}.

For completeness we present here the characteristic function $\omega_M(n, \gamma)$ of the kernel (\ref{KM}),
defined by the relation
\begin{equation}
\int d\vec{r}^{\;\prime}\langle\vec{r}|\hat{\mathcal{K}}_M|\vec{r^{\;\prime}}\rangle
e^{in\phi_{\vec r^{\;\prime}}}\left(\vec r^{\;\prime 2}\right)^{\gamma} =
\omega_M(n, \gamma)e^{in\phi_{\vec r}}\left(\vec r^{\;2}\right)^{\gamma}~.
\label{omega M}
\end{equation}
Actually because of running coupling the functions
$e^{in\phi_{\vec r^{\;\prime}}}\left(\vec r^{\;\prime 2}\right)^{\gamma}$  are not eigenfunctions of
$\hat{\mathcal{K}}_M$ anymore, and $\omega_M(n, \gamma)$ contains $\ln\vec r^{\;2}$.
In Eq.~(\ref{omega M}) it can be replaced by $\partial/\partial\gamma$. As the result we have
\[
\omega_M(n, \gamma) =\frac{\alpha_{s}(\mu^{2})N_{c}}{\pi}%
\chi(n,\gamma)+\frac{\alpha_{s}^{2}(\mu^{2})N_{c}^{2}}{4\pi^{2}}\left[
6\zeta\left(  3\right)  -2\Phi\left(  n,\gamma\right)  -2\Phi\left(
n,1-\gamma\right)  +F\left(  n,\gamma\right)  -\chi^{\prime\prime}\left(
n,\gamma\right)  \frac{{}}{{}}\right.
\]%
\begin{equation}
\left. + \left( \frac{67}{9}-2\zeta(2)\right)\chi(n,\gamma) +
\frac{11}{3}\left(\chi(n,\gamma)\left( \frac{\partial}{\partial\gamma
}-  \ln\left(\frac{4e^{-2C}}{\mu^2}\right)-\frac{2\gamma}{\gamma^{2}-\frac{n^{2}}{4}}\right)  +
\frac{\chi^{2}(n,\gamma)}{2}-\frac{\chi^{\prime}(n,\gamma
)}{2} \right) \right]  ,
\label{BFKL_in_gamma_repr}%
\end{equation}
where
\[
\Phi\left(  n,\gamma\right)  =\int_{0}^{1}\frac{dt}{1+t}t^{\gamma
-1+n/2}\left\{  \frac{\pi^{2}}{12}-\frac{1}{2}\psi^{\prime}\left(  \frac
{n+1}{2}\right)  -\operatorname{Li}_{2}\left(  t\right)  -\operatorname{Li}%
_{2}\left(  -t\right)  \right.
\]%
\begin{equation}
\left.  -\left(  \psi\left(  n+1\right)  +C+\ln\left(  1+t\right)  +\sum
_{k=1}^{\infty}\frac{\left(  -t\right)  ^{k}}{k+n}\right)  \ln t-\sum
_{k=1}^{\infty}\frac{t^{k}}{\left(  k+n\right)  ^{2}}\left[  1-\left(
-1\right)  ^{k}\right]  \right\}   \label{Phi1}%
\end{equation}
and%
\begin{equation}
F(n,\gamma)=\frac{\pi^{2}\cos\left(  \pi\gamma\right)  }{\sin^{2}\left(  \pi
\gamma\right)  \left(  1-2\gamma\right)  }\left(  \frac{\gamma\left(
1-\gamma\right)  \left(  \delta_{n,2}+\delta_{n,-2}\right)  }{2\left(
3-2\gamma\right)  \left(  1+2\gamma\right)  }-\left\{  \frac{3\gamma\left(
1-\gamma\right)  +2}{\left(  3-2\gamma\right)  \left(  1+2\gamma\right)
}+3\right\}  \delta_{n,0}\right)  . \label{F(n,gamma)}%
\end{equation}
The calculation of the integral (\ref{omega M}) is discussed in the Appendix C.

As it should be, the function $\omega_M(n, \gamma)$ differs from the corresponding function
$\omega(n, 1-\gamma)$  found in Ref.~\cite{Kotikov:2000pm} for the BFKL kernel in the momentum
representation  only by  the terms related with  renormalization (i.e. proportional to $\beta_0/N_c=11/3$ for
pure gluodynamics). Indeed, if the coupling was  not running,   $\omega_M(n, \gamma)$ and
$\omega(n, 1-\gamma)$ would be genuine eigenvalues corresponding to the same eigenstate  and should coincide.
Since $\omega_M(n, \gamma)$   was found using the results of
Refs.~\cite{Fadin:2006ha}, \cite{Fadin:2007ee} and \cite{Fadin:2007de} and the calculation of
$\omega(n, \gamma)$ in Ref.~\cite{Kotikov:2000pm} is based on the  results of  Ref.~\cite{FL98}, the
relationship between $\omega_M(n, \gamma)$ and  $\omega(n, \gamma)$ is the cross-check of all  these results.
The  coincidence  of the functions at $\beta_0=0$ gives a forcible argument in favour of their correctness.
Moreover, the terms proportional  to $\beta_0/N_c=11/3$ in $\omega_M(n, \gamma)$  can be derived from the
corresponding term in  $\omega(n, \gamma)$. Let the state $|n, \gamma\rangle$ be defined by the equality
$\langle\vec r|n, \gamma\rangle =e^{in\phi_{\vec r}}\left(\vec r^{\;2}\right)^{\gamma}$; then, using
\begin{equation}
\int_0^{2\pi} d\phi e^{in\phi +ia\cos\phi}=e^{in\pi/2}2\pi J_n(a)~, \;\;
\int_0^{\infty} dx x^\alpha J_n(b x)=2^\alpha b^{-\alpha-1}
\frac{\Gamma(\frac{n+1+\alpha}{2})}{\Gamma(\frac{n+1-\alpha}{2})}
 ~, \label{int}
\end{equation}%
where $J_n$ is the $n$-th Bessel function, we have
\begin{equation}
\langle\vec q|n, \gamma\rangle =e^{in\phi_{\vec q}}
\left(\vec q^{\; 2}\right)^{-1-\gamma}e^{i\pi n/2} \frac{2^{2\gamma+1}\Gamma(\frac{n}{2}+1+\gamma)}
{\Gamma(\frac{n}{2}-\gamma)}~.
\label{int-1}
\end{equation}%
The $\beta$-dependent terms in  $\omega(n, 1-\gamma)$ are written as
${\alpha_s^2N_c}\beta_0 R/{(4\pi^2)}$, where
\begin{equation}
R=-\ln\left( \frac{\vec q^{\;2}}{\mu^2}\right)\chi(n, \gamma)-\frac{\chi^2(n, \gamma)}{2}+
\frac{\chi^\prime(n, \gamma)}{2}
 ~.  \label{R definition}
\end{equation}
Since
\[
-\ln\left( \frac{\vec q^{\;2}}{\mu^2}\right)\chi(n, \gamma)\langle\vec q|n, \gamma\rangle =
\chi(n, \gamma)\left(\ln\mu^2+\frac{\partial}{\partial \gamma}
-\left[\frac{\partial}{\partial \gamma}\ln\left( \frac{2^{2\gamma+1}
\Gamma(\frac{n}{2}+1+\gamma)}{\Gamma(\frac{n}{2}-\gamma)}\right)\right]\right)\langle\vec q|n, \gamma\rangle\]
\begin{equation}
=\chi(n, \gamma)\left(\ln\mu^2+\frac{\partial}{\partial \gamma}
-2\ln 2+2C+\chi(n,\gamma)-\frac{2\gamma}{\gamma^2-\frac{n^2}{4}}\right)
\langle\vec q|n, \gamma\rangle ~,
\end{equation}%
and the M\"{o}bius form corresponds to the kernel obtained by the transformation (\ref{transform with beta}),
 which means
\begin{equation}
\omega(n, 1-\gamma)\rightarrow \omega(n, 1-\gamma) -\frac{\alpha_s^2N_c}{4\pi^2}\beta_0
\chi^\prime(n, \gamma)~,
\end{equation}
we come to the conclusion that in (\ref{omega M}) the $\beta$-dependent terms in
$\omega_M(n, \gamma)$ can be written in the form
\begin{equation}
\frac{\alpha_s^2N_c}{4\pi^2}\beta_0\left[\chi(n, \gamma)\left(\frac{\partial}{\partial \gamma} -
\ln\left(\frac{4e^{-2C}}{\mu^2}\right)
-\frac{2\gamma}{\gamma^2-\frac{n^2}{4}}\right)
+\frac{\chi^2(n, \gamma)}{2}-\frac{\chi^\prime(n, \gamma)}{2}
\right]~,
\end{equation}
which is exactly the same as in Eq.~(\ref{BFKL_in_gamma_repr}).

\section{SUSY Yang-Mills  forward kernel}

The extension of the NLO BFKL kernel to supersymmetric theories was performed
in Ref.~\cite{Kotikov:2000pm} for the forward case in the momentum representation
and in Ref.~\cite{Fadin:2007xy} for the nonforward case in the M\"{o}bius
coordinate representation.  Supersymmetric Yang-Mills theories contain gluons and $n_{M}$ Majorana
fermions in the adjoint representation of the color group. For $N$--extended
SUSY we have $n_{M}=N$. For $N>1$ besides fermions there are $n_{S}$ scalar particles;
$n_{S}=2$ at $N=2$ and $n_{S}=6$ at $N=4$. Following Ref.~\cite{Fadin:2007xy}
for the M\"{o}bius kernel in the SUSY theories we write %
\begin{equation}
g_{SUSY}=g_{Gluon}+g_{Fermion}+g_{Scalar}.
\end{equation}%
\begin{equation}
g_{SUSY}^{0}(\vec{r}_{1},\vec{r}_{2};\vec{\rho})=-g_{SUSY}(\vec{r}_{1},\vec
{r}_{2};\vec{\rho})+2\pi\zeta\left(  3\right)  \delta\left(  \vec{\rho
}\right)  ~, \label{g-three-point}%
\end{equation}%
\[
g_{SUSY}(\vec{r}_{1},\vec{r}_{2};\vec{r}_{2}^{\;\prime})\ =\frac{\vec{r}%
_{12}^{\;2}}{\vec{r}_{22^{\prime}}^{\;2}\vec{r}_{12^{\prime}}^{\;2}}\left[
\frac{67}{18}-\zeta(2)-\frac{5n_{M}+2n_{S}}{9}+\frac{\beta_{0}}{2N_{c}}%
\ln\left(  \frac{\vec{r}_{12}^{\;2}\mu^{2}}{4e^{-2C}}\right)  \right.
\]%
\begin{equation}
\left.  +\frac{\beta_{0}}{2N_{c}}\frac{\vec{r}_{12^{\prime}}^{\,\,\,2}-\vec
{r}_{22^{\prime}}^{\,\,\,2}}{\vec{r}_{12}^{\;2}}\ln\left(  \frac{\vec
{r}_{22^{\prime}}^{\,\,\,2}}{\vec{r}_{12^{\prime}}^{\,\,\,2}}\right)
-\frac{1}{2}\ln\left(  \frac{\vec{r}_{12}^{\;2}}{\vec{r}_{22^{\prime}}^{\;2}%
}\right)  \ln\left(  \frac{\vec{r}_{12}^{\;2}}{\vec{r}_{12^{\prime}}^{\;2}%
}\right)  +\frac{\vec{r}_{12^{\prime}}^{\;2}}{2\vec{r}_{12}^{\;2}}\ln\left(
\frac{\vec{r}_{12^{\prime}}^{\;2}}{\vec{r}_{22^{\prime}}^{\;2}}\right)
\ln\left(  \frac{\vec{r}_{12}^{\;2}}{\vec{r}_{12^{\prime}}^{\;2}}\right)
\right]  ~,
\end{equation}%
\[
g_{SUSY}(\vec{r}_{1},\vec{r}_{2};\vec{r}_{1}^{\;\prime},\vec{r}_{2}^{\;\prime
})=\frac{1}{2\vec{r}_{1^{\prime}2^{\prime}}^{\,\,4}}\left(  \frac{\vec
{r}_{11^{\prime}}^{\;2}\,\vec{r}_{22^{\prime}}^{\;2}-2\vec{r}_{12}^{\;2}%
\,\vec{r}_{1^{\prime}2^{\prime}}^{\;2}}{d}\ln\left(  \frac{\vec{r}%
_{12^{\prime}}^{\;2}\,\vec{r}_{21^{\prime}}^{\;2}}{\vec{r}_{11^{\prime}}%
^{\;2}\vec{r}_{22^{\prime}}^{\;2}}\right)  -1\right)  \left(  1-n_{M}%
+\frac{n_{S}}{2}\right)
\]%
\[
+\left(  \frac{(2n_{S}-3n_{M})}{4\vec{r}_{1^{\prime}2^{\prime}}^{\,\,2}}%
\frac{\vec{r}_{12}^{\;2}\,}{d}+\frac{1}{4\vec{r}\,\,_{11^{\prime}}^{2}%
\ \vec{r}\,\,_{22^{\prime}}^{2}}\left(  \frac{\vec{r}\,\,_{12}^{4}}{d}%
-\frac{\vec{r}\,\,_{12}^{2}}{\vec{r}\,\,_{1^{\prime}2^{\prime}}^{2}}\right)
\right)  \ln\left(  \frac{\vec{r}_{12^{\prime}}^{\;2}\,\vec{r}_{21^{\prime}%
}^{\;2}}{\vec{r}_{11^{\prime}}^{\;2}\vec{r}_{22^{\prime}}^{\;2}}\right)
\]%
\[
+\frac{\ln\ \left(  \frac{\vec{r}\,\,_{12}^{2}}{\vec{r}\,\,_{1^{\prime
}2^{\prime}}^{2}}\right)  }{4\vec{r}\,\,_{11^{\prime}}^{2}\vec{r}%
\,\,_{22^{\prime}}^{2}}\ +\frac{\ln\left(  \frac{\vec{r}\,\,_{12}^{2}\ \vec
{r}\,\,_{1^{\prime}2^{\prime}}^{2}}{\vec{r}\,\,_{11^{\prime}}^{2}\vec
{r}\,\,_{22^{\prime}}^{2}}\right)  }{2\vec{r}\,\,_{12^{\prime}}^{2}\vec
{r}\,\,_{21^{\prime}}^{2}\ }\left(  \frac{\vec{r}\,\,_{12}^{2}}{2\vec
{r}\,\,_{1^{\prime}2^{\prime}}^{2}}+\frac{1}{2}-\frac{\vec{r}\,\,_{22^{\prime
}}^{2}}{\vec{r}\,\,_{1^{\prime}2^{\prime}}^{2}}\right)  +\frac{\vec
{r}\,\,_{12}^{2}\ln\ \left(  \frac{\vec{r}\,\,_{12}^{2}\vec{r}\,\,_{1^{\prime
}2^{\prime}}^{2}}{\vec{r}\,\,_{12^{\prime}}^{2}\ \vec{r}\,\,_{21^{\prime}}%
^{2}}\right)  }{4\vec{r}\,\,_{11^{\prime}}^{2}\vec{r}\,\,_{22^{\prime}}%
^{2}\vec{r}\,\,_{1^{\prime}2^{\prime}}^{2}}%
\]%
\[
+\frac{\ln\left(  \frac{\vec{r}\,\,_{22^{\prime}}^{2}}{\vec{r}\,\,_{12}^{2}%
}\right)  }{2\vec{r}\,\,_{11^{\prime}}^{2}\ \vec{r}\,\,_{12^{\prime}}^{2}%
}+\frac{\ln\ \left(  \frac{\vec{r}\,\,_{12}^{2}\vec{r}\,\,_{1^{\prime
}2^{\prime}}^{2}}{\vec{r}\,\,_{12^{\prime}}^{2}\vec{r}\,\,_{22^{\prime}}^{2}%
}\right)  }{2\vec{r}\,\,_{11^{\prime}}^{2}\ \vec{r}\,\,_{1^{\prime}2^{\prime}%
}^{2}}+\frac{\ln\left(  \frac{\vec{r}\,\,_{12}^{2}\vec{r}\,\,_{11^{\prime}%
}^{2}}{\vec{r}\,\,_{22^{\prime}}^{2}\ \vec{r}\,\,_{1^{\prime}2^{\prime}}^{2}%
}\right)  }{2\vec{r}\,\,_{12^{\prime}}^{2}\vec{r}\,\,_{1^{\prime}2^{\prime}%
}^{2}}+\frac{\vec{r}\,\,_{12}^{2}\ln\ \left(  \frac{\vec{r}\,\,_{11^{\prime}%
}^{2}}{\vec{r}\,\,_{1^{\prime}2^{\prime}}^{2}}\right)  }{2\vec{r}%
\,\,_{11^{\prime}}^{2}\vec{r}\,\,_{12^{\prime}}^{2}\ \vec{r}\,\,_{22^{\prime}%
}^{2}}%
\]%
\begin{equation}
+\frac{\vec{r}\,\,_{21^{\prime}}^{2}\ln\left(  \frac{\vec{r}\,\,_{21^{\prime}%
}^{2}\vec{r}\,\,_{1^{\prime}2^{\prime}}^{2}}{\vec{r}\,\,_{12}^{2}\ \vec
{r}\,\,_{11^{\prime}}^{2}}\right)  }{2\vec{r}\,\,_{11^{\prime}}^{2}\vec
{r}\,\,_{22^{\prime}}^{2}\vec{r}\,\,_{1^{\prime}2^{\prime}}^{2}}%
+(1\leftrightarrow2,1^{\prime}\leftrightarrow2^{\prime}),\;\;d=\vec
{r}_{12^{\prime}}^{\;2}\vec{r}_{21^{\prime}}^{\;2}-\vec{r}_{11^{\prime}}%
^{\;2}\vec{r}_{22^{\prime}}^{\,\,2}. \label{g-four-point}%
\end{equation}
One should insert these functions $g$ into the definition of the M\"{o}bius
kernel in the coordinate representation (\ref{K_in_corrdinate_repr}).
Using the results of the previous section we can write the forward
M\"{o}bius kernel in the SUSY case. It reads
\[
\langle\vec{r}|\hat{{K}}^{SUSY}_M|\vec{r^{\;\prime}}\rangle
=\frac{\alpha_{s}(\frac{4e^{-2C}}{\vec{r}^{\:2}})N_{c}}{2\pi^2}\int
\frac{d\vec{\rho}\;\vec{r}^{\;2}}{(\vec{r}
-\vec{\rho})^{2}\vec{\rho}^{\:2}}\left\{
\left(2\delta(\vec{\rho}-\vec{r}^{\;\prime}) -\delta(\vec{r}-\vec{r^{\;\prime}})
\right)\left[1+\frac{\alpha_sN_c}{4\pi}
\left(\frac{67}{9}\right.\right.\right.
\]
\[
\left.\left.
-2\zeta(2)-\frac{10n_M}{9}-\frac{4n_S}{9}+\frac{\beta_0}{N_c}
\frac{\vec{\rho}^{\:2}-(\vec{r}
-\vec{\rho})^{2}}{\vec{r}^{\;2}}\ln\left(  \frac{(\vec{r}-\vec{\rho})^{2}}{\vec{\rho}^{\; 2}}\right)
\right]+ \frac{\alpha_sN_c}{4\pi}{3}\delta(\vec{r}-\vec{r^{\;\prime}})
\ln\left(  \frac{\vec{\rho}^{\:2}}{\vec{r}^{\;2}}\right)  \right.
\]
\begin{equation}
\left. \times \ln\left(  \frac{(\vec{r}-\vec{\rho})^2}{\vec{r}^{\;2}}\right)\right\}
+\frac{\alpha^2_{s}N^2_{c}}{4\pi^3}
\frac{\vec{r}^{\;2}}{\vec{r}^{\;\prime 2}}\left(f_1(\vec{r},\vec{r}^{\;\prime})
+f^{SUSY}_2(\vec{r},\vec{r}^{\;\prime}) -\frac{1}{(\vec{r}-\vec{r}^{\;\prime})^{2}}%
\ln^{2}\left(  \frac{\vec{r}^{\;2}}{\vec{r}^{\;\prime 2}}\right)\right) ~,
\label{KM SUSY coordinate}
\end{equation}
where
\begin{equation}
f^{SUSY}_2(\vec{r},\vec{r}^{\;\prime}) =(1-{n_M}+\frac{n_S}{2})f_2(\vec{r},\vec{r}^{\;\prime})+
(2n_S-3n_M)\int_0^\infty dt\frac{\ln\left|\frac{1+t}{1-t}\right|}{\vec{r}^{\;\prime\,2}+t^2\vec{r}^{\;2}}~,
\label{f2 susy}
\end{equation}
the functions $f_1$ and $f_2$ are defined in Eqs.~(\ref{f1}), (\ref{f2}),
\begin{equation}
\alpha_{s}(\frac{4e^{-2C}}{\vec{r}^{\:2}}) \simeq\alpha_{s}(\mu^{2})\left(  1-\frac{\alpha_{s}(\mu^{2}%
)}{4\pi}\beta_0\ln\left(  \frac{4e^{-2C}}{\vec{r}^{\:2}\mu^{2}}\right)
\right),\;\;\;\beta_{0}=\left(  \frac{11}{3}-\frac{2n_{M}}{3}-\frac{n_{S}}%
{6}\right)  N_{c}~,
\end{equation}
$\mu$ being the renormalization scale in the $\overline{MS}$-scheme.

As it is known,  at $N=4$ the coupling $\alpha_{s}$ does not run, so that $\beta
_{0}=0$. Moreover, it is seen from (\ref{f2 susy}) that $f^{SUSY}_2=0$ in this case. Next,  let us  express
our result in the renormalization scheme which preserves the supersymmetry. This scheme is known as the
dimensional reduction and it differs from the $\overline{MS}$-scheme in the finite charge renormalization
(see Ref.~\cite{Fadin:2007xy}) for  details). So we get
\begin{equation}
\alpha_{s}\rightarrow\alpha_{s}\left(  1-\frac{\alpha_{s}N_{c}}{12\pi}\right)
. \label{dim_reduction}%
\end{equation}
Finally, in the $N=4$ case, having $\beta_{0}=0,\,n_{S}=6,\,n_{M}=4$, the
kernel simplifies to%
\[
\langle\vec{r}|\hat{{K}}^{N=4}_M|\vec{r^{\;\prime}}\rangle
=\frac{\alpha_{s}N_{c}}{2\pi^2}\int
\frac{d\vec{\rho}\;\vec{r}^{\;2}}{(\vec{r}
-\vec{\rho})^{2}\vec{\rho}^{\:2}}
\left(2\delta(\vec{\rho}-\vec{r}^{\;\prime}) -\delta(\vec{r}-\vec{r^{\;\prime}})\right)
\left[1-\frac{\alpha_sN_c}{2\pi}\zeta(2)\right]
\]
\begin{equation}
+\frac{\alpha^2_{s}N^2_{c}}{4\pi^3}
\left[6\pi \zeta(3)\delta(\vec{r}-\vec{r^{\;\prime}})+\frac{\vec{r}^{\;2}}{\vec{r}^{\;\prime 2}}
\left(f_1(\vec{r},\vec{r}^{\;\prime})
 -\frac{1}{(\vec{r}-\vec{r}^{\;\prime})^{2}}%
\ln^{2}\left(  \frac{\vec{r}^{\;2}}{\vec{r}^{\;\prime 2}}\right)\right)\right] ~.
\label{K4 SUSY coordinate}
\end{equation}
Now let us consider the forward BFKL kernel in the momentum space. The explicit form of  the kernel can be
found for QCD in Ref.~\cite{FL98} and for Yang-Mills theory at N=4 in Ref.~\cite{Kotikov:2000pm}. In both
cases the kernel is written in the space-time dimension $D=4+2\epsilon$ to regularize infrared divergencies.
Here we solve two problems. First, we found the explicit form of  the kernel for SUSY Yang-Mills  with any N.
Second, we perform explicitly the cancellation of the infrared divergencies and write the kernel at physical
space-time dimension $D=4$.
It permits us to demonstrate  that the functional identity of the
forward BFKL kernels in the momentum and M\"{o}bius coordinate  representations exhibited in the previous
section in the LO   is preserved in the NLO in the  N=4 SUSY case.

To solve the first problem we must change the quark contribution in the kernel of
Ref.~\cite{FL98} for the gluino one and add the scalar particle contribution. It is known
(see Refs.~\cite{Kotikov:2000pm} and \cite{Fadin:2007xy})
that the gluino contribution to the``real" kernel can be
obtained from the quark one by the change of the coefficients, $n_f\rightarrow n_M N_c$ for the ``non-Abelian"
(leading at large $N_c$) part (including the trajectory) and $n_f\rightarrow -n_M N^3_c$ for the ``Abelian"
(suppressed at large $N_c$) part, so that this contribution can be found quite easy.  To obtain the scalar
contribution is a more subtle  task.  According to Ref.~\cite{Fadin:2007xy}, the scalar contribution also can
be divided into ``non-Abelian" and ``Abelian" parts.    Both in the integral representation of the  trajectory
and in the ``non-Abelian"  part  of the ``real" kernel the scalar contribution can be  obtained from the
corresponding quark contribution by the substitution $n_f\rightarrow n_SN_c$ and  the change of the fermion
polarization operator with the scalar one which differs from the former by the factor $1/(4(1+\epsilon))$.
We are interested in the kernel expanded in powers of $\epsilon$.  It is clear that since  the  factor
mentioned above depends on $\epsilon$  this kernel  can be  obtained applying substitutions, which are
different for the different terms  in  the expansion of the polarization operator. It is not difficult
to see (for details see Ref.~\cite{GF}) that these substitutions must be
\begin{equation}
\frac{2}{3}\frac{n_f}{N_c}\rightarrow \frac{n_S}{6},\;\;-\frac{10}{9}\frac{n_f}{N_c}
\rightarrow -\frac{4n_S}{9},\;\;\frac{56}{27}\frac{n_f}{N_c}\rightarrow \frac{26n_S}{27}.
\end{equation}
The $n$-th equality here corresponds to the $n$-th term in the expansion.

Therefore from Eq.~(6) of Ref.~\cite{FL98}  for the trajectory (in the $\overline{MS}$ scheme) we obtain
\[
\omega\left( - \vec{q}^{\,\,2}\right)  =-\bar{g}_{\mu}^{2}\left(  \frac
{2}{\epsilon}+2\ln\frac{\vec{q}^{\,\,2}}{\mu^{2}}\right)  \,-\bar{g}_{\mu
}^{4}\left[  \frac{\beta_0}{N_c}\left(  \frac{1}{\epsilon^{2}}-\ln{}^{2}\left(  \frac{\vec{q}^{\,\,2}}%
{\mu^{2}}\right)  \right)  \right.
\]%
\begin{equation}
\left.  +\left(  \frac{67}{9}-2\zeta\left(  2\right)  -\frac{10}{9}
n_{M}-\frac{4n_S}{9}\right)  \left(  \frac{1}{\epsilon}+2\ln\left(  \frac{\vec
{q}^{\,\,2}}{\mu^{2}}\right)  \right)  -\frac{404}{27}+2\zeta(3)+\frac
{56}{27}n_{M}+ \frac{26}{27}n_S\right]  ~,   \label{ew}%
\end{equation}%
where
\begin{equation}
g=g_{\mu}\mu^{-\mbox{\normalsize $\epsilon$}}\left[  1+\frac{\beta_0}{N_c}\frac{{\bar{g}}_{\mu
}^{2}}{2\epsilon} \right]  ~,\quad\quad{\bar{g}}_{\mu}^{2}=\frac{g^{2}_{\mu}N_{c}\Gamma(1-\epsilon
)}{\left(  4\pi\right)  ^{2+\epsilon}},\quad\quad\ \beta_0=
\left(  \frac{11}{3}-\frac{2}{3}n_{M}-\frac{n_S}{6}\right).
\end{equation}

The ``Abelian" part of the scalar contribution to the non-forward kernel is given by Eq.~(28) of
Ref.~\cite{Fadin:2007xy}. For the forward case, using the Feynman parametrization and performing integration
over $k_1$, for  $D=4$ (for details see Ref.~\cite{GF}) we obtain
\[
\langle\vec{q}|\hat{\mathcal{K}}^S_a|\vec{q}^{\,\,\prime}\rangle
=\frac{\alpha_s^2N_c^2}{4\pi^3}\frac{n_s}{2}\frac{1}
{16\vec{q}^{\,\,4}}\left\{\left(3(\vec{q}\vec{q}^{\,\,\prime})^2-2\vec{q}^{\,\,2}
\vec{q}^{\,\,\prime\,2}
\right)\left(\frac{2}{\vec{q}^{\,\,2}}+\frac{2}{\vec{q}^{\,\,\prime\,2}}
-\left(\frac{1}{\vec{q}^{\,\,2}}-\frac{1}{\vec{q}^{\,\,\prime\,2}}
\right)\right.\right.
\]
\[
\left.\left.\times\ln\left(\frac{\vec{q}^{\,\,2}}{\vec{q}^{\,\,\prime\,2}}\right)\right)
+\left[(\vec{q}\vec{q}^{\,\,\prime})^2\left(2-3\frac{\vec{q}^{\,\,\prime\,2}}
{\vec{q}^{\,\,2}}-3\frac{\vec{q}^{\,\,2}}{\vec{q}^{\,\,\prime\,2}}\right)+2\left(\vec{q}^{\,\,2}
+\vec{q}^{\,\,\prime\,2}\right)^2\right] \int_0^\infty
\frac {dt}{\vec{q}^{\,\,2}+t^2\vec{q}^{\,\,\prime\,2}}\ln\left|\frac{1+t}{1-t}
\right|\right\}
\]
\begin{equation}
=\frac{\alpha_s^2N_c^2}{4\pi^3}\frac{n_s}{2}
\frac{\vec{q}^{\,\,\prime\,2}}{\vec{q}^{\,\,2}}\left(f_2(\vec{q},\vec{q}^{\,\,\prime})
+4 \int_0^\infty \frac {dt}{\vec{q}^{\,\,2}+t^2\vec{q}^{\,\,\prime\,2}}\ln\left|\frac{1+t}{1-t}
\right|
\right)~,
\end{equation}
where $f_2$ is defined in Eq.~(\ref{f2}).  Taking into account that
$\langle\vec{q}|\hat{\mathcal{K}}|\vec{q}^{\,\,\prime}\rangle $ differs from
the symmetric kernel
\begin{equation}
K(\vec{q},\vec{q}^{\,\,\prime})=K_r\vec(\vec{q},\vec{q}^{\,\,\prime}) +
2\delta(\vec{q}-\vec{q}^{\,\,\prime})\omega\left( - \vec{q}^{\,\,2}\right)
\label{symmetric kernel}
\end{equation}
presented in Ref.~\cite{FL98} by the factor $\vec{q}^{\,\,\prime\,2}/\vec{q}^{\,\,2}$, for the ``real" part
we obtain
\[
K_r(\vec{q},\vec{q}^{\,\,\prime})
=\frac{4\,\overline{g}_{\mu}^{2}\,\mu^{-2\epsilon}}%
{\pi^{1+\epsilon}\Gamma(1-\epsilon)}\frac{1}%
{\vec{k}^{\,\,2}}+\frac{4\overline
{g}_{\mu}^{4}\,\mu^{-2\epsilon}}{\pi^{1+\epsilon}\Gamma(1-\epsilon)}
\]%
\[
\times\left\{\frac{1}{\,{\vec{k}^{\,\,2}}%
}\left[   \frac{\beta_0 }{N_c\epsilon
}\left(  1-\left(  \frac{\,{\vec{k}^{\,\,2}}}{\mu^{2}}\right)
^{\epsilon}(1-\epsilon^{2}\frac{\pi^{2}}{6})\right) +\left(  \frac{\,{\vec{k}^{\,\,2}}}{\mu^{2}%
}\right)  ^{\epsilon}\left(  \frac{67}{9}-\frac{\pi^{2}}{3}-\frac{10}{9}n_M
-\frac{4n_S}{9}\right. \right.  \right.  \,
\]%
\begin{equation}
\left.  \left. \left. +\epsilon\left(  -\frac{404}{27}+\frac{11}{3}\zeta(2)+14\zeta(3)+\frac{56}%
{27}n_{M} +\frac{26}{27}n_S \right)  -\ln^{2}\frac{\vec{q}^{\,\,2}}%
{\vec{q}^{\,\,\prime\,2}}\right)\right] +f_{1}\left(  \vec{q}_{1},\vec{q}_{1}^{\,\,\prime}\right)
+f_{2}^{SUSY}\left(  \vec{q}_{1},\vec{q}_{1}^{\,\,\prime}\right) \right\},  \label{eKr}%
\end{equation}
where $\vec k=\vec q-\vec q^{\;\prime}$, $f_1$ is defined in Eq.~(\ref{f2}) and $f_{2}^{SUSY}$ in
Eq.~(\ref{f2 susy}).

Eqs.~(\ref{ew}) and (\ref{eKr}) solve the first problem raised in the beginning of this section.
But they contain the infrared divergencies which complicate their use. One can cancel the divergencies
and write the kernel at physical space-time dimension $D=4$ following
Refs.~\cite{Fadin:2006ha}
and \cite{Fadin:2007de}
and using the integral representation for the trajectory
\begin{equation}
\omega(-\vec{q}^{\;2})=-\frac{\bar{g}_{\mu}^{2}\;\vec{q}^{\;2}}
{\pi^{1+\epsilon}\Gamma(1-\epsilon)}\int\frac{d^{2+2\epsilon}k\;\mu
^{-2\epsilon}}{\vec{k}^{\;2}(\vec{k}-\vec{q})^{2}}\left(  1+\bar{g}_{\mu
}^{2}f_{\omega}(\vec{k},\vec{k}-\vec{q})\right)  ~.
\label{omega as integral}
\end{equation}
The quark contribution to  $f_{\omega}$ is defined by Eqs.~(76) and (77) of Ref.~\cite{Fadin:2006ha}.
The gluino and scalar contributions can be obtained from the quark one by the substitutions discussed above.
The gluon contribution is given, with the required accuracy,
by Eq.~(23) of Ref.~\cite{Fadin:2007de}. As the result, we obtain with this accuracy   \[
f_{\omega}(\vec{k}_{1},\vec{k}_{2})=\frac{\beta_0}{N_c\epsilon}+\left[
\frac{\beta_0}{N_c\epsilon}-\frac{67}{9}+2\zeta\left(2\right)  +\frac{10}{9}
n_{M}+\frac{4n_S}{9}+\epsilon\left( \frac{404}{27}-\frac{11}{3}\zeta(2)-6\zeta(3)-\frac
{56}{27}n_{M}\right.\right.
\]
\begin{equation}
\left.\left.
-\frac{26}{27}n_S\right)\right]  \left[  \left(  \frac{\vec{k}_{12}^{\;2}}{\mu^{2}}\right)
^{\epsilon}-\left(  \frac{\vec{k}_{1}^{\;2}}{\mu^{2}}\right)  ^{\epsilon
}-\left(  \frac{\vec{k}_{2}^{\;2}}{\mu^{2}}\right)  ^{\epsilon}\right]
-\ln\left(  \frac{\vec{k}_{12}^{\;2}}{\vec{k}_{1}^{\;2}}\right)  \ln\left(
\frac{\vec{k}_{12}^{\;2}}{\vec{k}_{2}^{\;2}}\right)  ~.
\label{integrand for omega}
\end{equation}
Just like Refs. \cite{Fadin:2006ha} and \cite{Fadin:2007de}, in the limit $\epsilon\rightarrow0$  we  introduce
the cut-off $\lambda\rightarrow 0$ keeping  $\epsilon\ln\lambda\rightarrow 0$. Then in the
regions $\vec{k}^{\;2}\leq\lambda^{2}$ we have
\[
f_{\omega}(\vec{k},\vec{k}-\vec{q})=\frac{\beta_0}{N_c\epsilon}-\left(
\frac{\vec{k}^{\;2}}{\mu^{2}}\right)^{\epsilon}\left[
\frac{\beta_0}{N_c\epsilon}-\frac{67}{9}+2\zeta\left(2\right)  +\frac{10}{9}
n_{M}+\frac{4n_S}{9}\right.
\]
\begin{equation}
\left.
+\epsilon\left( \frac{404}{27}-\frac{11}{3}\zeta(2)-6\zeta(3)-\frac
{56}{27}n_{M}-\frac{26}{27}n_S\right)\right] \label{integrand for omega at small k}
\end{equation}
and in the region  $(\vec{k}-\vec{q})^{2}\leq\lambda^{2}$   the same expression with the substitution
$\vec{k}^{\;2}\rightarrow(\vec{k}-\vec{q})^{2}$.
Comparing Eq.~(\ref{integrand for omega at small k})  with Eq.~(\ref{eKr}) we  see that, when the kernel
$K(\vec{q},\vec{q}^{\,\,\prime})$ in Eq.~(\ref{symmetric kernel}) acts on any
function nonsingular at $\vec{q}=
\vec{q}^{\,\,\prime}$, the contribution of the region $\vec{k}^{\;2}\leq\lambda^{2}$ in
the ``real" part cancels almost completely the contributions of the regions $\vec{k}^{\;2}\leq\lambda^{2}$
and $(\vec{k}-\vec{q})^{2} \leq\lambda^{2}$ in the doubled trajectory $\omega(-\vec{q}^{\;2})$. The only
piece which remains uncancelled   is
\begin{equation}
2\frac{\bar{g}_{\mu}^{4}\;}{\pi^{1+\epsilon}\Gamma(1-\epsilon)}\int
\frac{d^{2+2\epsilon}k\;\mu^{-2\epsilon}}{\vec{k}^{\;2}}16\epsilon
\zeta(3)\left(  \frac{\vec{k}^{\;2}}{\mu^{2}}\right)  ^{\epsilon}
\theta(\lambda^{2}-\vec{k}^{\;2})=2\frac{\alpha_{s}^{2}(\mu)N_{c}^{2}}{2\pi
^{2}}\zeta(3). \label{uncancelled piece of omega}
\end{equation}
Outside the regions  $\vec{k}^{\;2}\leq\lambda^{2}$  and $(\vec{k}-\vec{q})^{2} \leq\lambda^{2}$ one  can
put $\epsilon=0$. Thus we come to the representation of the symmetric kernel
\[
K(\vec{q}, \vec{q^{\;\prime}})=
\frac{\alpha_{s}(\vec q^{\;2})N_{c}}{2\pi^{2}}\left[\frac{2}
{(\vec{q}-\vec{q}^{\;\prime})^{2}}-\delta(\vec{q}
-\vec{q}^{\;\prime})\int \frac{d\vec{l}\;\vec{q}^{\;2}}{(\vec{q}-\vec{l})^{2}
\vec{l}^{\:2}}\right]\left[1+\frac{\alpha_{s}N_{c}}{4\pi}\left(\frac{67}{9}-2\zeta\left(2\right)  -
\frac{10}{9}
n_{M}-\frac{4n_S}{9}\right)\right]
\]
\[
+\frac{\alpha^2_{s}N^2_{c}}{4\pi^{3}}\left[\frac{1}
{(\vec{q}-\vec{q}^{\;\prime})^{2}}\frac{\beta_0}{N_c}\ln\left(
\frac{\vec{q}^{\;2}}{(\vec{q}-\vec{q}^{\;\prime})^{\;2}}
\right)
+f_1(\vec{q}, \vec{q}^{\;\prime})+ f^{SUSY}_2(\vec{q}, \vec{q}^{\;\prime})-
\frac{1}{(\vec{q}-\vec{q}^{\;\prime})^2}\ln^2\left(  \frac{\vec{q}^{\;{2}}}{\vec{q}^{\;\prime 2}}\right)
\right.
\]
\begin{equation}
\left.+\delta(\vec{q}-\vec{q}^{\;\prime})\left(
\frac{\beta_0}{2N_c}\int \frac{d\vec{l}\;\vec{q}^{\;2}}{(\vec{q}-\vec{l})^{2}\vec{l}^{\:2}}
\ln\left(\frac{(\vec{q}-\vec{l})^{2}\vec{l}^{\:2}}{\vec{q}^{\;4}}
\right)+6\pi\zeta(3)\right)
\right]~,\label{SUSY symmetric kernel}
\end{equation}
which solves the second problem:  presentation of the kernel in the physical space-time dimension $D=4$
with  explicit cancellation of the infrared divergencies.

To compare the BFKL kernel in the M\"{o}bius representation (\ref{KM SUSY coordinate}) and in the momentum
representation, we have to take into account that the expression
 (\ref{KM SUSY coordinate}) corresponds to the kernel obtained from the symmetric
 one by the transformation (\ref{transform with beta}) with $\hat{\vec{q}}_{1}^{\;2}
 =\hat{\vec{q}}_{2}^{\;2}=\hat{\vec{q}}^{\;2}$  and that
$\langle\vec{q}|\hat{\mathcal{K}}|\vec{q}^{\;\prime}\rangle =
\vec{q}^{\;\prime\,2}K(\vec{q}, \vec{q}^{\;\prime})\vec{q}^{\;-2}$, so that
 \[
\langle\vec{q}|\hat{\mathcal{K}}|\vec{q}^{\;\prime}\rangle=
\frac{\alpha_{s}(\vec q^{\;2})N_{c}}{2\pi^{2}}\int
\frac{d\vec{l}\;\vec{q}^{\;\prime\,2}}{(\vec{q}-\vec{l})^{2}
\vec{l}^{\:2}}\left[2\delta(\vec{q}
-\vec{l})-\delta(\vec{q}
-\vec{q}^{\;\prime})\right]\left[1+\frac{\alpha_{s}N_{c}}{4\pi}\left(\frac{67}{9}
-2\zeta\left(2\right) \right.\right.
\]
\[
\left.\left. -\frac{10}{9}
n_{M}-\frac{4n_S}{9}\right)\right]+\frac{\alpha^2_{s}N^2_{c}}{4\pi^{3}}\frac{\vec{q}^{\;\prime\,2}}
{\vec{q}^{\;2}}\left[\frac{1}
{(\vec{q}-\vec{q}^{\;\prime})^{2}}\frac{\beta_0}{N_c}\ln\left(
\frac{\vec{q}^{\;2}}{(\vec{q}-\vec{q}^{\;\prime})^{\;2}}
\right)
+f_1(\vec{q}, \vec{q}^{\;\prime})+ f^{SUSY}_2(\vec{q}, \vec{q}^{\;\prime})
\right.
\]
\begin{equation}
\left.-
\frac{1}{(\vec{q}-\vec{q}^{\;\prime})^2}\ln^2\left(  \frac{\vec{q}^{\;{2}}}{\vec{q}^{\;\prime 2}}\right)+\delta(\vec{q}-\vec{q}^{\;\prime})\left(
\frac{\beta_0}{2N_c}\int \frac{d\vec{l}\;\vec{q}^{\;2}}{(\vec{q}-\vec{l})^{2}\vec{l}^{\:2}}
\ln\left(\frac{(\vec{q}-\vec{l})^{2}\vec{l}^{\:2}}{\vec{q}^{\;4}}
\right)+6\pi\zeta(3)\right)
\right]~.\label{SUSY  kernel}
\end{equation}
Comparing this expression with Eq.~(\ref{KM SUSY coordinate})  one can see
that at $\beta=0$ they are functionally identical up to the normalization factors:
\begin{equation}
\frac{\vec{r}^{\,\,\prime\,2}}{\vec{r}^{\,\,2}}\langle\vec{r}|\hat{\mathcal{K}}%
_{M}|\vec{r}^{\,\,2}\rangle|_{\beta_0=0}=\frac{\vec{q}^{\;2}}{\vec{q}^{\;\prime2}%
}\langle\vec{q}|\hat{\mathcal{K}}|\vec{q}^{\,\,\prime}%
\rangle_{\beta_0=0}\Bigg|_{\vec{q}\rightarrow\vec{r},\,\,\vec{q}^{\,\,\prime
}\rightarrow\vec{r}^{\,\,\prime
}}~. \label{functional_identity}%
\end{equation}
Note that  with these normalization factors the kernels are symmetric
with respect to $\vec{r}\leftrightarrow\vec{r}^{\,\,\prime}$\ or $\vec{q}_{1}%
\leftrightarrow\vec{q}_{1}^{\,\,\prime}$ \ substitution.

Actually the identity relation can be expected, because  at $\beta=0$ the kernels
$\hat{\vec{r}}^{\,\,-2}\hat{\mathcal{K}}_M\hat{\vec{r}}^{\,\,2}$ and
$\hat{\vec{q}}^{\,\,2}\hat{\mathcal{K}}\hat{\vec{q}}^{\,\,-2}$ have the same eigenvalues corresponding
to the eigenfunctions connected by the replacement $\vec{r}\leftrightarrow \vec{q}$.

\section{Comparison of the kernels: nonforward case}

In this section  we present the results of the simplest generalizations for
the nonforward case of
the transformation  used in section 4 for the elimination of the discrepancy
between the results of Refs.~\cite{Fadin:2007de} and \cite{Balitsky:2008zz}
for the forward scattering.   We  start from  the commutator
$\left[  \mathcal{\hat{K}}%
^{(B)},\,\mathcal{\hat{K}}^{(B)}\frac{1}{4}\ln\left(  \hat{q}_{1}^{2}\hat{q}%
_{2}^{2}\right)\right]  $.  The calculation of this  commutator
is described in   the appendix B. The result is
\[
\left(  \frac{2\pi^{2}}{\alpha_{s}N_{c}}\right)  ^{2}\langle\vec{r}_{1}%
,\vec{r}_{2}|\left[  \mathcal{\hat{K}}^{(B)},\,\mathcal{\hat{K}}^{(B)}\frac{1}%
{4}\ln\left(  \hat{q}_{1}^{2}\hat{q}_{2}^{2}\right)\right]_{M}  |\vec{r}%
_{1}^{\,\prime},\vec{r}_{2}^{\,\prime\;}\rangle
\]%
\[
=\frac{1}{2}\int d\vec{\rho}\frac{\vec{r}_{12}^{\,\,2}}{\vec{r}_{1^{\prime
}2^{\prime}}^{\,\,2}\vec{r}_{1\rho}^{\,\,2}\vec{r}_{2\rho}^{\,\,2}}\left(
\left(  \delta\left(  \vec{r}_{11^{\prime}}\right)  -\delta\left(  \vec
{r}_{1^{\prime}\rho}\right)  \right)  \frac{(\vec{r}_{21^{\prime}}\,\vec
{r}_{1^{\prime}2^{\prime}})}{\vec{r}_{22^{\prime}}^{\,\,2}}\ln\left(
\frac{\vec{r}_{21^{\prime}}^{\,\,2}}{\vec{r}_{1^{\prime}2^{\prime}}^{\,\,2}%
}\right)  +\delta\left(  \vec{r}_{11^{\prime}}\right)  \frac{(\vec
{r}_{1^{\prime}2^{\prime}}\,\vec{r}_{1^{\prime}\rho})}{\vec{r}_{2^{\prime}%
\rho}^{\,\,2}}\ln\left(  \frac{\vec{r}_{1^{\prime}\rho}^{\,\,2}}{\vec
{r}_{1^{\prime}2^{\prime}}^{\,\,2}}\right)  \right)
\]%
\begin{equation}
+\vec{r}_{12}^{\,\,2}\left(  \frac{\ln\left(  \frac{\vec{r}_{12^{\prime}%
}^{\,\,2}\vec{r}_{21^{\prime}}^{\,\,2}}{\vec{r}_{12}^{\,\,2}\vec{r}%
_{1^{\prime}2^{\prime}}^{\,\,2}}\right)  }{4\vec{r}_{11^{\prime}}^{\,\,2}%
\vec{r}_{22^{\prime}}^{\,\,2}\vec{r}_{1^{\prime}2^{\prime}}^{\,\,2}}+\frac
{\ln\left(  \frac{\vec{r}_{11^{\prime}}^{\,\,2}\vec{r}_{22^{\prime}}^{\,\,2}%
}{\vec{r}_{12}^{\,\,2}\vec{r}_{1^{\prime}2^{\prime}}^{\,\,2}}\right)  }%
{4\vec{r}_{11^{\prime}}^{\,\,2}\vec{r}_{21^{\prime}}^{\,\,2}}\left(  \frac
{1}{\vec{r}_{1^{\prime}2^{\prime}}^{\,\,2}}-\frac{1}{\vec{r}_{22^{\prime}%
}^{\,\,2}}\right)  \right)  +\frac{\ln\left(  \frac{\vec{r}_{11^{\prime}%
}^{\,\,2}\vec{r}_{22^{\prime}}^{\,\,2}}{\vec{r}_{12^{\prime}}^{\,\,2}\vec
{r}_{21^{\prime}}^{\,\,2}}\right)  }{4\vec{r}_{11^{\prime}}^{\,\,2}\vec
{r}_{1^{\prime}2^{\prime}}^{\,\,2}}\left(  \frac{\vec{r}_{12^{\prime}}%
^{\,\,2}}{\vec{r}_{22^{\prime}}^{\,\,2}}-1\right)  +(1\leftrightarrow
2,1^{\prime}\leftrightarrow2^{\prime}).\label{[K,KL]}
\end{equation}
The subscript ${M}$ means that the operator acts in the M\"{o}bius representation,  i.e.
the matrix element (\ref{[K,KL]}) vanishes as $\vec{r}_{1}\rightarrow\vec{r}_{2}.$
One can see that the above
expression has ultraviolet  singularities  which cancel in  the convolution with
M\"{o}bius impact factors. But the structure of these singularities is
different from the structure of singularities in the gluon contribution to
the kernel given in Eqs.~(\ref{g0BFKL}) and
(\ref{g3BFKL}). Therefore it is obvious that this commutator cannot make
$g^{0}$ and $g(\vec{r}_{1},\vec{r}_{2},\vec{r}_{2}^{\,\,\prime})$ coincident
with the corresponding $g_{BC}^{0}$ and $g_{BC}^{0}(\vec{r}_{1},\vec{r}%
_{2},\vec{r}_{2}^{\,\,\prime}).$ As for $g(\vec{r}_{1},\vec{r}_{2};\vec{r}%
_{1}^{\;\prime},\vec{r}_{2}^{\;\prime}),$ we have
\begin{align}
&  g(\vec{r}_{1},\vec{r}_{2};\vec{r}_{1}^{\;\prime},\vec{r}_{2}^{\;\prime
})-g_{BC}(\vec{r}_{1},\vec{r}_{2};\vec{r}_{1}^{\;\prime},\vec{r}_{2}%
^{\;\prime})=\frac{\ln\ \left(  \frac{\vec{r}\,\,_{12}^{2}}{\vec
{r}\,\,_{1^{\prime}2^{\prime}}^{2}}\right)  }{4\vec{r}\,\,_{11^{\prime}}%
^{2}\vec{r}\,\,_{22^{\prime}}^{2}}+\frac{\ln\ \left(  \frac{\vec{r}%
\,\,_{12}^{2}\vec{r}\,\,_{1^{\prime}2^{\prime}}^{2}}{\vec{r}\,\,_{11^{\prime}%
}^{2}\vec{r}\,\,_{22^{\prime}}^{2}}\right)  }{4\vec{r}\,\,_{12^{\prime}}%
^{2}\ \vec{r}\,\,_{21^{\prime}}^{2}}-\frac{\ln\left(  \frac{\vec
{r}_{12^{\prime}}^{\;2}\,\vec{r}_{21^{\prime}}^{\;2}}{\vec{r}_{11^{\prime}%
}^{\;2}\vec{r}_{22^{\prime}}^{\;2}}\right)  }{4\vec{r}_{1^{\prime}2^{\prime}%
}^{\,\,4}}\nonumber\\
&  +\frac{\ln\left(  \frac{\vec{r}\,\,_{22^{\prime}}^{2}}{\vec{r}\,\,_{12}%
^{2}}\right)  }{2\vec{r}\,\,_{11^{\prime}}^{2}\ \vec{r}\,\,_{12^{\prime}}^{2}%
}+\frac{\ln\ \left(  \frac{\vec{r}\,\,_{12}^{2}\vec{r}\,\,_{1^{\prime
}2^{\prime}}^{2}}{\vec{r}\,\,_{12^{\prime}}^{2}\vec{r}\,\,_{22^{\prime}}^{2}%
}\right)  }{2\vec{r}\,\,_{11^{\prime}}^{2}\ \vec{r}\,\,_{1^{\prime}2^{\prime}%
}^{2}}+\frac{\ln\left(  \frac{\vec{r}\,\,_{12}^{2}\vec{r}\,\,_{11^{\prime}%
}^{2}}{\vec{r}\,\,_{22^{\prime}}^{2}\ \vec{r}\,\,_{1^{\prime}2^{\prime}}^{2}%
}\right)  }{2\vec{r}\,\,_{12^{\prime}}^{2}\vec{r}\,\,_{1^{\prime}2^{\prime}%
}^{2}}+\frac{\vec{r}\,\,_{22^{\prime}}^{2}\ln\left(  \frac{\vec{r}%
\,\,_{11^{\prime}}^{2}\vec{r}\,\,_{22^{\prime}}^{2}}{\vec{r}\,\,_{12}%
^{2}\ \vec{r}\,\,_{1^{\prime}2^{\prime}}^{2}}\right)  }{2\vec{r}%
\,\,_{12^{\prime}}^{2}\vec{r}\,\,_{21^{\prime}}^{2}\vec{r}\,\,_{1^{\prime
}2^{\prime}}^{2}}+\frac{\vec{r}\,\,_{21^{\prime}}^{2}\ln\left(  \frac{\vec
{r}\,\,_{21^{\prime}}^{2}\vec{r}\,\,_{1^{\prime}2^{\prime}}^{2}}{\vec
{r}\,\,_{12}^{2}\ \vec{r}\,\,_{11^{\prime}}^{2}}\right)  }{2\vec
{r}\,\,_{11^{\prime}}^{2}\vec{r}\,\,_{22^{\prime}}^{2}\vec{r}\,\,_{1^{\prime
}2^{\prime}}^{2}}\nonumber\\
&  +\frac{\vec{r}\,\,_{12}^{2}\ln\ \left(  \frac{\vec{r}\,\,_{11^{\prime}}%
^{2}}{\vec{r}\,\,_{1^{\prime}2^{\prime}}^{2}}\right)  }{2\vec{r}%
\,\,_{11^{\prime}}^{2}\vec{r}\,\,_{12^{\prime}}^{2}\ \vec{r}\,\,_{22^{\prime}%
}^{2}}+\frac{\vec{r}\,\,_{12}^{2}\ln\ \left(  \frac{\vec{r}\,\,_{12}^{2}%
\vec{r}\,\,_{1^{\prime}2^{\prime}}^{2}}{\vec{r}\,\,_{12^{\prime}}^{2}\ \vec
{r}\,\,_{21^{\prime}}^{2}}\right)  }{4\vec{r}\,\,_{11^{\prime}}^{2}\vec
{r}\,\,_{22^{\prime}}^{2}\vec{r}\,\,_{1^{\prime}2^{\prime}}^{2}}+\frac{\vec
{r}\,\,_{12}^{2}\ln\left(  \frac{\vec{r}\,\,_{12}^{2}\ \vec{r}\,\,_{1^{\prime
}2^{\prime}}^{2}}{\vec{r}\,\,_{11^{\prime}}^{2}\vec{r}\,\,_{22^{\prime}}^{2}%
}\right)  }{4\vec{r}\,\,_{12^{\prime}}^{2}\vec{r}\,\,_{21^{\prime}}^{2}%
\ \vec{r}\,\,_{1^{\prime}2^{\prime}}^{2}}+(1\leftrightarrow2,1^{\prime
}\leftrightarrow2^{\prime})
\end{align}
and again we see that this commutator does not help to eliminate the discrepancy.

Then one can search for a commutator equal to $\left[  \mathcal{\hat{K}}%
^{(B)},\,\mathcal{\hat{K}}^{(B)}\frac{1}{4}\ln\left(  \hat{q}_{1}^{2}\hat{q}%
_{2}^{2}\right) \right]_M  $ in the forward case but different from it in
general. An example of this kind is $\left[  \mathcal{\hat{K}}^{(B)},\,\frac
{1}{4}\ln\left(  \hat{q}_{1}^{2}\hat{q}_{2}^{2}\right)  \mathcal{\hat{K}}%
^{B}\right]  _{M}.$ Indeed, this commutator coincides with the previous one in the forward case
since acting on the  eigenfunctions
$\langle \vec q|n, \gamma \rangle \propto e^{in\phi_{\vec q}}\left(\vec q^{\;2}\right)^{\gamma-2}$
they yield the same result:
\begin{align}
\langle\vec{q}|\left[  \mathcal{\hat{K}}^{(B)},\,\mathcal{\hat{K}}^{(B)}%
\ln\left(  \hat{q}^{2}\right)  \right]  |n, \gamma\rangle &  =\int d^{2}%
l\,\langle\vec{q}|\mathcal{\hat{K}}^{(B)}\left[  \mathcal{\hat
{K}}^{(B)},\ln\left(  \hat{q}^{2}\right)  \right]  |\vec{l}\rangle\langle\vec{l}|n, \gamma\rangle\nonumber\\
&  =\chi\left( n, \gamma\right)  \chi^{\prime}\left( n, \gamma\right)  \langle
\vec{q}|n, \gamma\rangle = \langle\vec{q}|\left[  \mathcal{\hat{K}}%
^{(B)},\,\ln\left(  \hat{q}^{2}\right)  \mathcal{\hat{K}}^{(B)}\right]
|n, \gamma\rangle.
\end{align}
Here  we have used the equality
$(\vec{q}^{\,\,2})^{\gamma}\ln\left(  \vec{q}^{\,\,2}\right)  =
{\partial(\vec{q}^{\,\,2})^{\gamma}}/(\partial\gamma)$\ to perform the
integration. The coordinate representation of $\left[  \mathcal{\hat{K}}%
^{(B)},\,\frac{1}{4}\ln\left(  \hat{q}_{1}^{2}\hat{q}_{2}^{2}\right)
\mathcal{\hat{K}}^{(B)}\right]  _{M}$ is also given in the appendix B and it
also fails to resolve the difference between the BFKL and BK kernels.
Thus, the simplest  generalizations are unable to eliminate the discrepancy in the nonforward case.

\section{Summary}

In this paper we  studied the properties of the  next-to-leading   BFKL kernel  in gluodynamics  and SUSY
Yang-Mills theories. In particular our study is connected with  the discrepancy  between the gluon
contribution to  the BFKL kernel in the M\"{o}bius representation found in Ref. \cite{Fadin:2007de} and
the kernel of the linearized BK equation calculated in Ref.~\cite{Balitsky:2008zz}.

We analyzed the ambiguity of the next-to-leading kernel, in particular in
its connection  with the energy scale.
Broadly speaking, the ambiguity  is  related to the rearrangement of  the radiative corrections between the
kernel and the impact factors, and the one connected  with the energy scale is not an exception. It is
shown that the change of the energy scale  can be associated with the specific form of the general
transformation of the NLO kernel discussed in Refs.~\cite{Fadin:2006ha} and
\cite{Fadin:2007de}.

The ambiguity can be used to remove the discrepancy  between the results of Refs. \cite{Fadin:2007de}
and ~\cite{Balitsky:2008zz}. It was explicitly demonstrated in the case of forward scattering. We found
the M\"{o}bius kernel for this case and showed that the major part of the difference between this kernel
and  the corresponding kernel of Ref.~\cite{Balitsky:2008zz} can be eliminated by the suitable
transformormation which can be associated with the change of the energy scale, so that the difference is
reduced to two terms. One of them is proportional to  the first coefficient of the $\beta$-function.
In our opinion, this term  is connected with the difference of  the renormalization scheme used in
Ref.~\cite{Balitsky:2008zz} from the conventional $\overline{MS}$-scheme. This term can be eliminated by
the change of the  renormalization scheme.  Unfortunately, we cannot  eliminate the third term,
proportional to $\zeta(3)$. In the BFKL approach this term passed through a great number of verifications.
It is also confirmed by the calculation of the three-loop anomalous dimensions in
Refs.~\cite{Vogt:2004mw} and \cite{{Kotikov:2004er}}.\footnote{When this paper was completed,  the article \cite{Balitsky:2009xg} appeared in the web. It is stated  in this article that the discussed  term appeared in the difference   because of the erroneous calculation of the integral in Ref.~\cite{Balitsky:2008zz}.}

We calculated the characteristic function  $\omega_M(n, \gamma)$ describing
the action of the forward
M\"{o}bius kernel on the eigenfunctions of the leading order kernel  and compared it with the corresponding
function of the BFKL kernel in the momentum representation found in Ref.~\cite{Kotikov:2000pm}.
The forward M\"{o}bius kernel was found using the results of
Refs.~\cite{Fadin:2006ha}, \cite{Fadin:2007ee} and \cite{ Fadin:2007de}
whereas the calculation of Ref.~\cite{Kotikov:2000pm}
is based on the  results of  Ref.~\cite{FL98}. Therefore the comparison  serves as the cross-check of
the results of  these papers.   The coincidence of the characteristic functions gives a strong  argument
in favour of rightness of the used results.

We studied also the forward BFKL kernel in  supersymmetric Yang-Mills theories
for any N-extended SUSY  both in the momentum and in the M\"{o}bius coordinate representations and
demonstrated the functional identity (up to the normalization factors) of the form of the kernel in
these representations for $N=4$.  We calculated  the  kernel in the M\"{o}bius representation in the impact
parameter space using the results of Refs.~\cite{Fadin:2006ha},
\cite{Fadin:2007ee}, \cite{ Fadin:2007de} and\cite{ Fadin:2007xy},  and
found the kernel for any N in the momentum space using the results of Refs.~\cite{FL98} and
\cite{Kotikov:2000pm}.
Performing explicit  cancellation of the infrared divergencies and writing the kernel at physical space-time
dimension $D=4$ we demonstrated the functional identity mentioned above,
that confirms the correctness of the used result.

At last, we checked how the simplest generalizations of the transformation,  used for the elimination of the
discrepancy between the results of Refs.~\cite{Fadin:2007de} and \cite{Balitsky:2008zz} for the forward
scattering, work in the general (nonforward) case. Unfortunately, these generalizations are not effective. This does not mean that the transformation eliminating the discrepancy (apart from the difference
in the renormalization scheme and in the $\zeta(3)$ term) does not exist. Moreover, we hope that it exists;
but in this case the generalization from the forward case is more refined than that we used.

\vspace{1.0cm} \noindent {\large \bf Acknowledgment}
\vspace{0.5cm}

One of us (V.S.F.) thanks L.N. Lipatov for numerous discussions, the
Alexander von Humboldt foundation for the re-invitation, the Dipartimento di Fisica dell'Universit\`a della
Calabria and the Istituto Nazionale di Fisica Nucleare, Gruppo
Collegato di Cosenza, the
Universit\"at Hamburg and DESY for their warm hospitality while a
part of this work was done.

\newpage

\section*{Appendix A}
In the representation  (\ref{forward_kernel_coord})
for  $\langle\vec{r}|\hat{\mathcal{K}}|\vec{r}^{\;\prime}\rangle$  only the term with
$g(\vec{r}_{1},\vec{r}_{2};\vec{r}%
_{1}^{\;\prime},\vec{r}_{2}^{\;\prime})$ in
$\langle\vec{r}_{1},\vec{r}_{2}|\hat{\mathcal{K}}|\vec{r}_1^{\;\prime}\vec{r}_2
^{\;\prime}\rangle$ requires integration.  Let us introduce
\begin{equation}
L(\vec{x},\vec{z})=\frac{1}{\pi}\int g(\vec{r}_{1},\vec{r}_{2};\vec{r}%
_{1}^{\;\prime},\vec{r}_{2}^{\;\prime})\delta
(\vec{r}_{1^{\prime}2^{\prime}}-\vec{z})d^{\,\,2}r_{1}^{\;\prime}d^{\,\,2}r_{2}^{\;\prime}%
\end{equation}
where $\vec{r}_{12} =\vec{x}$ and $g(\vec{r}_{1},\vec{r}_{2};\vec{r}_{1}^{\,\,2},\vec
{r}_{2}^{\,\,2})$ is defined in Eq.~(\ref{gBFKL}). Denoting
\begin{equation}
f_{1}\left(  \vec{x},\vec{z}\right)  =\frac{\vec{z}^{\,\,2}}{\vec{x}^{\,\,2}}\int
\frac{d\vec{z}_{1}}{2\pi}\left[  \frac{\vec{x}^{4}}{(\vec{x}-\vec{z}-\vec
{z}_{1})^{\,\,2}\vec{z}_{1}^{\,\,2}-(\vec{x}-\vec{z}_{1})^{\,\,2}(\vec{z}+\vec{z}_{1}%
)^{\,\,2}}+\frac{\vec{x}^{\,\,2}}{\vec{z}^{\,\,2}}\right]  \frac{\ln\left(  \frac{(\vec
{x}-\vec{z}-\vec{z}_{1})^{\,\,2}\vec{z}_{1}^{\,\,2}}{(\vec{x}-\vec{z}_{1})^{\,\,2}(\vec
{z}+\vec{z}_{1})^{\,\,2}}\right)  }{\vec{z}_{1}^{\,\,2}(\vec{x}-\vec{z}-\vec{z}%
_{1})^{\,\,2}}%
~\label{f1 definition}%
\end{equation}
and
\begin{equation}
f_{2}\left(  \vec{x},\vec{z}\right)  =\frac{2}{\vec{x}^{\,\,2}\vec{z}^{\,\,2}}%
\int\frac{d^{\,\,2}\vec{z}_{1}}{2\pi}\left(  \frac{\left(  (\vec{x}-\vec{z}%
-\vec{z}_{1})^{\,\,2}\vec{z}_{1}^{\,\,2}-2\vec{x}^{\,\,2}\vec{z}^{\,\,2}\right)  \ln\left(
\frac{(\vec{x}-\vec{z}_{1})^{\,\,2}(\vec{z}+\vec{z}_{1})^{\,\,2}}{(\vec{x}-\vec
{z}-\vec{z}_{1})^{\,\,2}\vec{z}_{1}^{\,\,2}}\right)  }{(\vec{x}-\vec{z}_{1})^{\,\,2}%
(\vec{z}+\vec{z}_{1})^{\,\,2}-(\vec{x}-\vec{z}-\vec{z}_{1})^{\,\,2}\vec{z}_{1}^{\,\,2}%
}-1\right)~, \label{f2 definition}
\end{equation}
we have
\[
L\left(  \vec{x},\vec{z}\right)  =\frac{\vec{x}^{\,\,2}}{\vec{z}^{\,\,2}}f_{1}\left(
\vec{z},\vec{x}\right)  +\frac{\vec{x}^{\,\,2}}{\vec{z}^{\,\,2}}f_{2}\left(  \vec
{x},\vec{z}\right)
\]%
\[
+\frac{1}{2\pi}\int d\vec{z}_{1}\left[  \frac{\ln\left(  \frac{\vec{x}^{\,\,2}%
}{\vec{z}^{\,\,2}}\right)  }{2(\vec{x}-\vec{z}-\vec{z}_{1})^{\,\,2}\vec{z}_{1}^{\,\,2}%
}+\frac{\ln\left(  \frac{(\vec{x}-\vec{z}-\vec{z}_{1})^{\,\,2}\vec{z}_{1}^{\,\,2}%
}{\vec{x}^{\,\,2}\vec{z}^{\,\,2}}\right)  }{(\vec{x}-\vec{z}_{1})^{\,\,2}(\vec{z}+\vec
{z}_{1})^{\,\,2}}\left(  \frac{\vec{z}_{1}^{\,\,2}}{\vec{z}^{\,\,2}}-\frac{\vec{x}^{\,\,2}%
}{2\vec{z}^{\,\,2}}-\frac{1}{2}\right)  +\frac{\ln\left(  \frac{\vec{x}^{\,\,2}\vec
{z}^{\,\,2}}{(\vec{x}-\vec{z}_{1})^{\,\,2}(\vec{z}+\vec{z}_{1})^{\,\,2}}\right)  \vec
{x}^{\,\,2}}{2\vec{z}^{\,\,2}(\vec{x}-\vec{z}-\vec{z}_{1})^{\,\,2}\vec{z}_{1}^{\,\,2}}\right.
\]%
\[
+\frac{\ln\left(  \frac{\vec{x}^{\,\,2}(\vec{x}-\vec{z}-\vec{z}_{1})^{\,\,2}}{\vec
{z}^{\,\,2}\vec{z}_{1}^{\,\,2}}\right)  }{\vec{z}^{\,\,2}(\vec{x}-\vec{z}_{1})^{\,\,2}}%
+\frac{\ln\left(  \frac{\vec{z}_{1}^{\,\,2}}{\vec{x}^{\,\,2}}\right)  }{(\vec{x}%
-\vec{z}_{1})^{\,\,2}(\vec{x}-\vec{z}-\vec{z}_{1})^{\,\,2}}+\frac{\ln\left(
\frac{(\vec{x}-\vec{z}-\vec{z}_{1})^{\,\,2}}{\vec{z}^{\,\,2}}\right)  \vec{x}^{\,\,2}%
}{(\vec{x}-\vec{z}_{1})^{\,\,2}(\vec{x}-\vec{z}-\vec{z}_{1})^{\,\,2}\vec{z}_{1}^{\,\,2}}%
\]%
\begin{equation}
\left.  +\frac{\ln\left(  \frac{\vec{z}^{\,\,2}(\vec{z}+\vec{z}_{1})^{\,\,2}}{\vec
{x}^{\,\,2}(\vec{x}-\vec{z}-\vec{z}_{1})^{\,\,2}}\right)  (\vec{z}+\vec{z}_{1})^{\,\,2}%
}{\vec{z}^{\,\,2}(\vec{x}-\vec{z}-\vec{z}_{1})^{\,\,2}\vec{z}_{1}^{\,\,2}}+\frac
{\ln\left(  \frac{\vec{x}^{\,\,2}\vec{z}^{\,\,2}}{(\vec{x}-\vec{z}_{1})^{\,\,2}\vec{z}%
_{1}^{\,\,2}}\right)  }{\vec{z}^{\,\,2}(\vec{x}-\vec{z}-\vec{z}_{1})^{\,\,2}}+(\vec{z}%
_{1}\rightarrow\vec{x}-\vec{z}-\vec{z}_{1})\right]  .\label{integral in L}
\end{equation}
The function $f_{1}$ can be obtained from the integral $J_{13}$ in
\cite{Fadin:199} and \cite{Kotsky:1998ug}. It reads
\[
f_{1}\left(  \vec{x},\vec{z}\right)  =\frac{\vec{z}^{\,\,2}}{\vec{x}^{\,\,2}}\int
\frac{d\vec{z}_{1}}{2\pi}\left[  \frac{\vec{x}^{4}}{(\vec{x}-\vec{z}-\vec
{z}_{1})^{\,\,2}\vec{z}_{1}^{\,\,2}-(\vec{x}-\vec{z}_{1})^{\,\,2}(\vec{z}+\vec{z}_{1}%
)^{\,\,2}}+\frac{\vec{x}^{\,\,2}}{\vec{z}^{\,\,2}}\right]  \frac{\ln\left(  \frac{(\vec
{x}-\vec{z}-\vec{z}_{1})^{\,\,2}\vec{z}_{1}^{\,\,2}}{(\vec{x}-\vec{z}_{1})^{\,\,2}(\vec
{z}+\vec{z}_{1})^{\,\,2}}\right)  }{\vec{z}_{1}^{\,\,2}(\vec{x}-\vec{z}-\vec{z}%
_{1})^{\,\,2}}%
\]%
\[
=\frac{\left(  \vec{x}^{\,\,2}-\vec{z}^{\,\,2}\right)  }{\left(  \vec{x}-\vec
{z}\right)  ^{\,\,2}\left(  \vec{x}+\vec{z}\right)  ^{\,\,2}}\left[  \ln\left(
\frac{\vec{x}^{\,\,2}}{\vec{z}^{\,\,2}}\right)  \ln\left(  \frac{\vec{x}^{\,\,2}\vec
{z}^{\,\,2}\left(  \vec{x}-\vec{z}\right)  ^{4}}{\left(  \vec{x}^{\,\,2}+\vec{z}%
^{\,\,2}\right)  ^{4}}\right)  +2\operatorname{Li}_{2}\left(  -\frac{\vec{z}^{\,\,2}%
}{\vec{x}^{\,\,2}}\right)  -2\operatorname{Li}_{2}\left(  -\frac{\vec{x}^{\,\,2}}%
{\vec{z}^{\,\,2}}\right)  \right]
\]%
\begin{equation}
-\left(  1-\frac{\left(  \vec{x}^{\,\,2}-\vec{z}^{\,\,2}\right)  ^{\,\,2}}{\left(  \vec
{x}-\vec{z}\right)  ^{\,\,2}\left(  \vec{x}+\vec{z}\right)  ^{\,\,2}}\right)  \left[
\int_{0}^{1}-\int_{1}^{\infty}\right]  \frac{du}{\left(  \vec{x}-\vec
{z}u\right)  ^{\,\,2}}\ln\left(  \frac{u^{\,\,2}\vec{z}^{\,\,2}}{\vec{x}^{\,\,2}}\right)
~.\label{f1-KLF}%
\end{equation}
To find  $f_{2}$ we use the representation
\begin{equation}
\frac{\ln\left(  \frac{ab}{cd}\right)  }{ab-cd}=\int_{0}^{1}\frac
{du}{((1-u)c+ua)((1-u)b+ud)}~,%
\end{equation}
which permits to integrate over  $\vec{z}_{1}$ using the Feynman
parametrization and the dimensional regularization
and gives $f_{2}$ in the form
\begin{equation}
f_{2}(\vec{x},\vec{z})=4\left(  \frac{\left(  \vec{x}\hspace{0.25em}\vec
{z}\right)  ^{\,\,2}}{\vec{x}^{\,\,2}\vec{z}^{\,\,2}}+2\right)  J_{2}-2J_{\vec{x}%
}-2J_{\vec{z}}-\frac{3}{2}J_{1}.
\end{equation}
Here
\[
J_{1}=\int\frac{d^{2+2\epsilon}\vec{z}_{1}}{\pi^{1+\epsilon}\Gamma\left(
1-\epsilon\right)  }\frac{\ln\left(  \frac{(\vec{x}-\vec{z}_{1})^{\,\,2}(\vec
{z}+\vec{z}_{1})^{\,\,2}}{(\vec{x}-\vec{z}-\vec{z}_{1})^{\,\,2}\vec{z}_{1}^{\,\,2}%
}\right)  }{(\vec{x}-\vec{z}_{1})^{\,\,2}(\vec{z}+\vec{z}_{1})^{\,\,2}-(\vec{x}%
-\vec{z}-\vec{z}_{1})^{\,\,2}\vec{z}_{1}^{\,\,2}}%
\]%
\begin{equation}
=\int_{0}^{1}du\int_{0}^{1}dv\frac{1}{u(1-u)\vec{z}^{\hspace{0.25em}%
2}+v(1-v)\vec{x}^{\hspace{0.25em}\hspace{0.25em}2}}=-2\int_{0}^{\infty}%
dt\frac{\ln\left\vert \frac{1-t}{1+t}\right\vert }{\vec{z}^{\,\,2}+t^{\,\,2}\vec
{x}^{\,\,2}},\label{j1}%
\end{equation}%
\[
J_{\vec{x}}=\int_{0}^{1}du\int_{0}^{1}dv\frac{v(1-v)}{u(1-u)\vec{z}%
^{\hspace{0.25em}2}+v(1-v)\vec{x}^{\hspace{0.25em}\hspace{0.25em}2}}
\]%
\begin{equation}
=-\frac
{1}{4}\left(  \left(  1-\frac{\vec{z}^{\,\,2}}{\vec{x}^{\,\,2}}\right)  \int
_{0}^{\infty}dt\frac{\ln\left\vert \frac{1-t}{1+t}\right\vert }{\vec{z}%
^{\,\,2}+t^{\,\,2}\vec{x}^{\,\,2}}-\frac{2}{\vec{x}^{\,\,2}}+\frac{\ln\frac{\vec{x}^{\,\,2}}%
{\vec{z}^{\,\,2}}}{\vec{x}^{\,\,2}}\right)  ,
\end{equation}%
\[
J_{2}=\int_{0}^{1}du\int_{0}^{1}dv\frac{u(1-u)v(1-v)}{u(1-u)\vec{z}%
^{\hspace{0.25em}2}+v(1-v)\vec{x}^{\hspace{0.25em}\hspace{0.25em}2}}
\]%
\begin{equation}
=-\frac{1}{32}\left[  \left(  1-\frac{3}{2}\left(  \frac{\vec{z}^{\,\,2}}{\vec{x}^{\,\,2}%
}+\frac{\vec{x}^{\,\,2}}{\vec{z}^{\,\,2}}\right)  \right)  \int_{0}^{\infty}%
dt\frac{\ln\left\vert \frac{1-t}{1+t}\right\vert }{\vec{z}^{\,\,2}+t^{\,\,2}\vec
{x}^{\,\,2}}\right. \left.  -3\left(  \frac{1}{\vec{x}^{\,\,2}}+\frac{1}{\vec{z}^{\,\,2}}\right)
+\frac{3}{2}\ln\frac{\vec{x}^{\,\,2}}{\vec{z}^{\,\,2}}\left(  \frac{1}{\vec{x}^{\,\,2}%
}-\frac{1}{\vec{z}^{\,\,2}}\right)  \right]  ,
\end{equation}
and $J_{\vec{z}}$ can be obtained from $J_{\vec{x}}$ via $\vec{x}%
\leftrightarrow\vec{z}$ substitution. These integrals can be calculated using their analytical properties.
Let us  consider $J_{1}$  and write it as
\begin{equation}
J_{1}(\vec{x},\vec{z})=\frac{1}{\vec{z}^{\,\,2}}f(t)|_{t=-\frac{\vec{x}^{\,\,2}}%
{\vec{z}^{\,\,2}}},\quad f(t)=\int_{0}^{1}du\int_{0}^{1}dv\frac{1}{u(1-u)-v(1-v)t}%
.
\end{equation}
Finding  the imaginary part of $f(t+i0)$
\begin{equation}
\operatorname{Im}f(t+i0)=\pi\int_{0}^{1}du\int_{0}^{1}dv\delta
(u(1-u)-v(1-v)t)=\frac{1}{\sqrt{t}}\ln\left\vert \frac{1+t}{1-t}\right\vert
\theta(t)
\end{equation}
and  restoring $f(t)$ as
\begin{equation}
f(t)=\frac{1}{\pi}\int_{0}^{\infty}\frac{\operatorname{Im}f(t^{\prime}%
+i0)}{t^{\prime}-t}dt^{\prime}.\label{caushi}%
\end{equation}
we arrive to Eq.~(\ref{j1}). The other functions $J$ were calculated in the
similar way. Finally we get $f_{2}(\vec{x},\vec{z})$ in the form (\ref{f2}).

Although the whole integral in Eq.~(\ref{integral in L}) converges,
the separate terms diverge, so that we use
dimensional regularization to calculate them. We need the following integrals:
\begin{align}
\int\frac{d^{2+2\epsilon}l}{\pi^{1+\epsilon}\Gamma\left(  1-\epsilon\right)
}\frac{1}{l^{2}(l+p)^{2}}  &  =\frac{2\left(  \ln\left(  p^{2}\right)
+\frac{1}{\epsilon}\right)  }{p^{2}},\\
\int\frac{d^{2+2\epsilon}l}{\pi^{1+\epsilon}\Gamma\left(  1-\epsilon\right)
}\frac{\ln\left(  l^{2}\right)  }{(l+p)^{2}}  &  =\frac{1}{2}\ln^{2}\left(
p^{2}\right)  +\frac{\ln\left(  p^{2}\right)  }{\epsilon}+\frac{1}%
{\epsilon^{2}}-\frac{\pi^{2}}{6},
\end{align}%
\begin{align}
\int\frac{d^{2+2\epsilon}l}{\pi^{1+\epsilon}\Gamma\left(  1-\epsilon\right)
}\frac{\ln\left(  \frac{l^{2}}{\mu^{2}}\right)  }{l^{2}(l+p)^{2}}  &
=-\frac{\frac{1}{2}\ln^{2}\left(  p^{2}\right)  +\frac{\ln\left(
p^{2}\right)  }{\epsilon}+\frac{1}{\epsilon^{2}}-\frac{\pi^{2}}{6}}{p^{2}%
}-\frac{2\left(  \ln\left(  p^{2}\right)  +\frac{1}{\epsilon}\right)
\ln\left(  \frac{\mu^{2}}{p^{2}}\right)  }{p^{2}},\\
\int\frac{d^{2+2\epsilon}l}{\pi^{1+\epsilon}\Gamma\left(  1-\epsilon\right)
}\frac{\ln\left(  \frac{l^{2}}{\mu^{2}}\right)  }{(l+p)^{2}(l+\mu)^{2}}  &
=\frac{\left(  \ln(p-\mu)^{2}+\frac{1}{2}\ln\left(  \frac{p^{2}}{\mu^{2}%
}\right)  +\frac{1}{\epsilon}\right)  \ln\left(  \frac{p^{2}}{\mu^{2}}\right)
}{(p-\mu)^{2}},
\end{align}%
\[
\int\frac{d^{2+2\epsilon}l}{\pi^{1+\epsilon}\Gamma\left(  1-\epsilon\right)
}\frac{\ln\left(  \frac{l^{2}}{\mu^{2}}\right)  }{l^{2}(l+p)^{2}(l-q)^{2}%
}=\frac{-\frac{1}{2}\ln^{2}\left(  p^{2}\right)  -\frac{\ln\left(
p^{2}\right)  }{\epsilon}-\frac{1}{\epsilon^{2}}+\frac{\pi^{2}}{6}}{p^{2}%
q^{2}}+\frac{\left(  \ln\left(  q^{2}\right)  +\frac{1}{\epsilon}\right)
\ln\left(  \frac{q^{2}}{\mu^{2}}\right)  }{q^{2}(p+q)^{2}}%
\]%
\[
-\frac{\ln\left(  \frac{(p+q)^{2}}{q^{2}}\right)  \ln\left(  \frac{p^{2}q^{2}%
}{\mu^{4}}\right)  \left(  (p+q)^{2}-q^{2}\right)  }{2p^{2}q^{2}(p+q)^{2}%
}-\frac{2I\left(  p^{2},q^{2},(p+q)^{2}\right)  \left(  p^{2}q^{2}%
-(p\,q)^{2}\right)  }{p^{2}q^{2}(p+q)^{2}}%
\]%
\begin{equation}
+\frac{\left(  \ln\left(  p^{2}\right)  +\frac{1}{\epsilon}\right)  \ln\left(
\frac{p^{2}}{\mu^{2}}\right)  }{p^{2}}\left(  \frac{1}{q^{2}}+\frac
{1}{(p+q)^{2}}\right)  +\frac{\ln\left(  \frac{(p+q)^{2}}{p^{2}}\right)
\ln\left(  \frac{p^{2}q^{2}}{\mu^{4}}\right)  }{2q^{2}(p+q)^{2}},
\label{91}
\end{equation}%
\[
\int\frac{d^{2+2\epsilon}l}{\pi^{1+\epsilon}\Gamma\left(  1-\epsilon\right)
}\frac{\ln\left(  \frac{l^{2}}{\mu^{2}}\right)  (l+q)^{2}}{(l+p)^{2}%
(l+\mu)^{2}}=\frac{1}{2}\ln^{2}\left(  \mu^{2}\right)  +\frac{\ln\left(
\mu^{2}\right)  }{\epsilon}+\frac{1}{\epsilon^{2}}-\frac{\pi^{2}}{6}%
+\frac{(q-p)(p-\mu)\ln^{2}\left(  \frac{p^{2}}{\mu^{2}}\right)  }{2(p-\mu
)^{2}}%
\]%
\[
-\frac{\ln\left(  \frac{p^{2}\mu^{2}}{(p-\mu)^{4}}\right)  (q-p)(p-\mu
)\ln\left(  \frac{p^{2}}{\mu^{2}}\right)  }{2(p-\mu)^{2}}+\frac{\left(
\ln(p-\mu)^{2}+\frac{1}{2}\ln\left(  \frac{p^{2}}{\mu^{2}}\right)  +\frac
{1}{\epsilon}\right)  \ln\left(  \frac{p^{2}}{\mu^{2}}\right)  (q-p)^{2}%
}{(p-\mu)^{2}}%
\]%
\begin{equation}
+2I\left(  p^{2},\mu^{2},(p-\mu)^{2}\right)  \left(  \frac{p(q-p)\left(
\left(  p\,\mu\right)  -\mu^{2}\right)  }{(p-\mu)^{2}}+\frac{\left(
(p\,\mu)-p^{2}\right)  (q-p)\mu}{(p-\mu)^{2}}\right)  ,
\label{92}
\end{equation}%
\[
\int\frac{d^{2+2\epsilon}l}{\pi^{1+\epsilon}\Gamma\left(  1-\epsilon\right)
}\frac{\ln\left(  \frac{l^{2}}{\mu^{2}}\right)  (l+q)^{2}}{l^{2}(l+p)^{2}%
}=\frac{\left(  \frac{1}{2}\ln^{2}\left(  p^{2}\right)  +\frac{\ln\left(
p^{2}\right)  }{\epsilon}+\frac{1}{\epsilon^{2}}-\frac{\pi^{2}}{6}\right)
\left(  p^{2}-q^{2}\right)  }{p^{2}}%
\]%
\begin{equation}
-\frac{\left(  \ln\left(  p^{2}\right)  +\frac{1}{\epsilon}\right)  \ln\left(
\frac{\mu^{2}}{p^{2}}\right)  \left(  q^{2}-p^{2}+(p-q)^{2}\right)  }{p^{2}}.
\end{equation}
The function $I$, which appears in Eqs.~(\ref{91}) and (\ref{92}), is given by
\begin{equation}
I(q^{2},p^{2},\mu^{2})=\int_{0}^{1}\frac{dx}{q^{2}(1-x)+p^{2}x-\mu^{2}%
x(1-x)}\ln\left(  \frac{q^{2}(1-x)+p^{2}x}{\mu^{2}x(1-x)}\right)  ~.
\label{integral I}%
\end{equation}
Using the integrals presented above we obtain
\begin{equation}
L\left(  x,z\right)  =\frac{x^{2}}{z^{2}}f_{1}\left(  z,x\right)  +\frac
{x^{2}}{z^{2}}f_{2}\left(  x,z\right)  +\ln\left(  \frac{x^{2}}{z^{2}}\right)
\ln\left(  \frac{(x-z)^{2}}{z^{2}}\right)  \left(  \frac{1}{(x-z)^{2}}%
-\frac{\,x^{2}}{(x-z)^{2}z^{2}}\right)  .
\end{equation}
Adding the contributions of the functions $g^0(\vec{r}_{1},\vec{r}_{2};\vec{\rho})$
and $g(\vec{r}_{1},\vec{r}_{2};\vec{\rho})$  we arrive to Eq.~(\ref{KM}).

\section*{Appendix B}

Here we will describe the calculation of the commutators. The commutator
necessary to eliminate the energy scale dependent terms in the difference of
the forward kernels (\ref{K M - K BC}) is $\left[  \mathcal{\hat{K}}^{(B)}%
\,,\ln\left(  \hat{\vec{q}}^{\,\,2}\right)  \mathcal{\hat{K}}^{(B)}\right]  .$
We will calculate it in the momentum space via the identity
\begin{equation}
\langle\vec{q}|\left[  \mathcal{\hat{K}}^{(B)},\,\ln\left(  \hat{\vec{q}%
}^{\,\,2}\right)  \mathcal{\hat{K}}^{(B)}\right]  |\vec{q}^{\,\,\prime}%
\rangle=\int d\vec{p}\langle\vec{q}|\left[  \mathcal{\hat{K}}^{(B)},\,\ln\left(
\hat{\vec{q}}^{\,\,2}\right)  \right]  |\vec{p}\rangle\langle\vec
{p}|\mathcal{\hat{K}}^{(B)}|\vec{q}^{\,\,\prime}\rangle.
\label{commutator_forward_mom}%
\end{equation}
Taking the LO forward kernel from Eq.~(\ref{LO BFKL_forward_kernel})\ we get%
\begin{equation}
\langle\vec{q}|\left[  \mathcal{\hat{K}}^{(B)},\,\ln\left(  \hat{\vec{q}%
}^{\,\,2}\right)  \right]  |\vec{p}\rangle=\frac{\alpha_{s}N_{c}}{\pi^{2}%
}\frac{\vec{p}^{\,2}}{(\vec{p}-\vec{q})^{2}\vec{q}^{\,\,2}}\ln\left(
\frac{\vec{p}^{\,\,2}}{\vec{q}^{\,\,2}}\right)  .
\end{equation}
Then for the whole commutator (\ref{commutator_forward_mom}) in the momentum
representation we have
\begin{equation}
\langle\vec{q}|\left[  \mathcal{\hat{K}}^{(B)},\,\ln\left(  \hat{\vec{q}%
}^{\,\,2}\right)  \mathcal{\hat{K}}^{(B)}\right]  |\vec{q}^{\,\prime}%
\rangle=\frac{\alpha_{s}^{2}N_{c}^{2}}{2\pi^{3}}\frac{\vec{q}^{\,\,\prime2}%
}{\vec{q}^{\,\,2}(\vec{q}-\vec{q}^{\,\,\prime})^{2}}\ln\left(  \frac{(\vec
{q}-\vec{q}^{\,\,\prime})^{4}}{\vec{q}^{\,\,2}\vec{q}^{\,\,\prime2}}\right)
\ln\left(  \frac{\vec{q}{}^{\,\,\prime2}}{\vec{q}^{\,\,2}}\right)  .
\label{commutator_forward}%
\end{equation}
Now we will rewrite this result in the coordinate space. Since we need the
operator in the M\"{o}bius representation, i.e.  with the matrix element equal to 0 at $\vec
{r}=0$, we should Fourier transform this expression and subtract from it its
value at $\vec{r}=0.$\ This subtraction allows us to cancel the singularity
at $\vec{q}=0$ in  Eq.~(\ref{commutator_forward}). To find the Fourier transform it
is convenient to rewrite  Eq.~(\ref{commutator_forward}) as%
\[
\langle\vec{q}|\left[  \mathcal{\hat{K}}^{(B)},\,\ln\left(  \hat{\vec{q}%
}^{\,\,2}\right)  \mathcal{\hat{K}}^{(B)}\right]  |\vec{q}^{\,\prime}%
\rangle=\frac{\alpha_{s}^{2}N_{c}^{2}}{2\pi^{3}}\left(  \frac{1}{\vec
{q}^{\,\,2}}+\frac{1}{(\vec{q}-\vec{q}^{\,\,\prime})^{2}}-\frac{2\vec{q}%
(\vec{q}-\vec{q}^{\,\,\prime})}{\vec{q}^{\,\,2}(\vec{q}-\vec{q}^{\,\,\prime
})^{2}}\right)
\]%
\begin{equation}
\times\left(  \ln^{2}\left(  \frac{(\vec{q}-\vec{q}^{\,\,\prime})^{2}}{\vec
{q}^{\,\,2}}\right)  -\ln^{2}\left(  \frac{(\vec{q}-\vec{q}^{\,\,\prime})^{2}%
}{\vec{q}{}^{\,\,\prime2}}\right)  \right)
\end{equation}
and use the following integrals:
\begin{equation}
\int\frac{d\vec{q}}{2\pi}\int\frac{d\vec{k}}{2\pi}e^{i[\vec{q}\,\vec{r}%
+\vec{k}\,\vec{\rho}]}\frac{1}{\vec{q}^{\,2}}\ln^{2}\left(  \frac{(\vec
{q}+\vec{k})^{2}}{\vec{k}^{\,2}}\right)  =\frac{1}{\vec{\rho}^{\,2}}\ln
^{2}\left(  \frac{\left(  \vec{r}-\vec{\rho}\right)  ^{2}}{\vec{r}^{\,2}%
}\right)  ,
\end{equation}%
\begin{equation}
\int\frac{d\vec{q}}{2\pi}\int\frac{d\vec{k}}{2\pi}e^{i[\vec{q}\,\vec{r}%
+\vec{k}\,\vec{\rho}]}\frac{(\vec{q}\,\vec{k})}{\vec{q}^{\,2}\vec{k}^{\,2}}%
\ln^{2}\left(  \frac{(\vec{k}+\vec{q})^{2}}{\vec{q}^{\,2}}\right)
=-\frac{\left(  \vec{r}\,\vec{\rho}\right)  }{\vec{r}^{\,2}\vec{\rho}^{\,2}%
}\ln^{2}\left(  \frac{(\vec{\rho}-\vec{r})^{2}}{\vec{\rho}^{\,2}}\right)  ,
\end{equation}%
\begin{equation}
\int\frac{d\vec{q}}{2\pi}\int\frac{d\vec{k}}{2\pi}e^{i[\vec{q}\,\vec{r}%
+\vec{k}\,\vec{\rho}]}\frac{(\vec{q}\,\vec{k})}{\vec{q}^{\,2}\vec{k}^{\,2}}%
\ln^{2}\left(  \frac{\vec{k}^{\,2}}{\vec{q}^{\,2}}\right)  =-\frac{\left(
\vec{r}\,\vec{\rho}\right)  }{\vec{r}^{\,2}\vec{\rho}^{\,2}}\ln^{2}\left(
\frac{\vec{\rho}^{\,2}}{\vec{r}^{\,2}}\right)  ,
\end{equation}%
\[
\int\frac{d\vec{q}}{2\pi}\int\frac{d\vec{k}}{2\pi}\frac{1}{\vec{k}^{\,2}}%
\ln^{2}\left(  \frac{\vec{k}^{\,2}}{\vec{q}^{\,2}}\right)  \left(
e^{i[\vec{q}\,\vec{r}+\vec{k}\,\vec{\rho}]}-e^{i[\vec{q}\,\vec{r}+\vec
{k}\,(\vec{\rho}-\vec{r})]}\right)  
\]%
\begin{equation}
=\frac{1}{\vec{r}^{\,2}}\left(  \ln
^{2}\left(  \frac{\vec{\rho}^{\,2}}{\vec{r}^{\,2}}\right)  -\ln^{2}\left(
\frac{(\vec{\rho}-\vec{r})^{2}}{\vec{r}^{\,2}}\right)  \right)  ,
\end{equation}%
\begin{equation}
\int\frac{d\vec{q}}{2\pi}\int\frac{d\vec{k}}{2\pi}\frac{1}{\vec{q}^{\,2}}%
\ln^{2}\left(  \frac{\vec{k}^{\,2}}{\vec{q}^{\,2}}\right)  \left(
e^{i[\vec{q}\,(\vec{r}-\vec{\rho})-\vec{k}\,\vec{\rho}]}-e^{-i[\vec{q}%
\,+\vec{k}]\,\vec{\rho}}\right)  =\frac{1}{\vec{\rho}^{\,2}}\ln^{2}\left(
\frac{(\vec{\rho}-\vec{r})^{2}}{\vec{\rho}^{\,2}}\right)  .
\end{equation}
As the result, we obtain
\[
\langle\vec{r}|\left[  \mathcal{\hat{K}}^{(B)},\,\ln\left(  \hat{\vec{q}%
}^{\,\,2}\right)  \mathcal{\hat{K}}^{(B)}\right]  _{M}|\vec{r}^{\,\prime}%
\rangle=\int\frac{d\vec{q}}{2\pi}\frac{d\vec{q}^{\,\,\prime}}{2\pi}\langle
\vec{q}|\left[  \mathcal{\hat{K}}^{(B)},\,\ln\left(  \hat{\vec{q}}%
^{\,\,2}\right)  \mathcal{\hat{K}}^{(B)}\right]  |\vec{q}^{\,\prime}%
\rangle\left(  e^{i\,\vec{q}\,\vec{r}-i\,\vec{q}\,^{\,\prime}\vec
{r}^{\,\,\prime}}-e^{-i\,\vec{q}\,^{\,\prime}\vec{r}^{\,\,\prime}}\right)
\]%
\begin{equation}
=-\frac{\alpha_{s}^{2}N_{c}^{2}}{2\pi^{3}}\frac{\,\vec{r}^{\,\,2}}{(\vec
{r}-\vec{r}^{\prime})^{2}\vec{r}^{\,\,\prime2}}\ln\left(  \frac{\vec
{r}^{\,\,2}}{\vec{r}^{\,\,\prime2}}\right)  \ln\left(  \frac{\vec{r}%
^{\,\,2}\vec{r}^{\,\,\prime2}}{(\vec{r}-\vec{r}^{\prime})^{4}}\right)  ,
\label{commutator_forward_coord}%
\end{equation}
\ which eliminates a part of the difference between the kernels in
Eq.~(\ref{K M - K BC}).

A natural generalization of the previous commutator to the nonforward case is
\begin{equation}
\left[  \mathcal{\hat{K}}^{(B)},\,\ln\left(  \hat{\vec{q}}_{1}^{\,\,2}\hat
{\vec{q}}_{2}^{\,\,2}\right)  \mathcal{\hat{K}}^{(B)}\right]  .
\end{equation}
We will also calculate it in the momentum representation and then
Fourier transform it to the coordinate space. In the momentum representation
we have%
\[
\langle\vec{q}_{1},\vec{q}_{2}|\left[  \mathcal{\hat{K}}^{(B)},\,\ln\left(
\hat{\vec{q}}_{1}^{\,\,2}\hat{\vec{q}}_{2}^{\,\,2}\right)  \mathcal{\hat{K}%
}^{(B)}\right]  |\vec{q}_{1}^{\,\prime},\vec{q}_{2}^{\,\prime\;}\rangle
\]%
\[
=\int d\vec{p}_{1}d\vec{p}_{2}\langle\vec{q}_{1},\vec{q}_{2}|\left[  \mathcal{\hat
{K}}^{(B)}\mathcal{\,}\ln\left(  \hat{\vec{q}}_{1}^{\,\,2}\hat{\vec{q}}%
_{2}^{\,\,2}\right)  \right]  |\vec{p}_{1},\vec{p}_{2}\rangle\langle\vec
{p}_{1},\vec{p}_{2}|\mathcal{\hat{K}}^{(B)}|\vec{q}_{1}^{\,\prime},\vec{q}%
_{2}^{\,\prime\;}\rangle
\]%
\[
=\delta\left(  \vec{q}-\vec{q}^{\,\prime}\right)  \int d\vec{k}_{1}%
\frac{\mathcal{K}_{r}^{(B)}(\vec{q}_{1},\vec{q}_{1}-\vec{k}_{1},\vec
{q})\mathcal{K}_{r}^{(B)}(\vec{q}_{1}-\vec{k}_{1},\vec{q}_{1}^{\,\,\prime}%
,\vec{q}^{\,\,\prime})}{\vec{q}_{1}^{\,\,2}\vec{q}_{2}^{\,\,2}(\vec{q}%
_{1}-\vec{k}_{1})^{\,\,2}(\vec{q}_{2}+\vec{k}_{1})^{\,\,2}}\ln\left(
\frac{(\vec{q}_{1}-\vec{k}_{1})^{\,\,2}(\vec{q}_{2}+\vec{k}_{1})^{\,\,2}}%
{\vec{q}_{1}^{\,\,2}\vec{q}_{2}^{\,\,2}}\right)
\]%
\begin{equation}
+\delta\left(  \vec{q}-\vec{q}^{\,\prime}\right)  \frac{\mathcal{K}_{r}%
^{(B)}(\vec{q}_{1},\vec{q}_{1}^{\,\,\prime},\vec{q})}{\vec{q}_{1}^{\,\,2}\vec
{q}_{2}^{\,\,2}}\left(  \omega\left(  \vec{q}_{1}^{\,\prime2}\right)
+\omega\left(  \vec{q}_{2}^{\,\prime2}\right)  \right)  \ln\left(  \frac
{\vec{q}_{1}^{\,\prime2}\vec{q}_{2}^{\,\prime2}}{\vec{q}_{1}^{\,\,2}\vec
{q}_{2}^{\,\,2}}\right)  .
\end{equation}
It is more convenient for the integration to rewrite the commutator in the
following form%
\[
\left(  \frac{\pi^{\,\,2}}{\alpha N}\right)  ^{\,\,2}\langle\vec{r}_{1}%
,\vec{r}_{2}|\left[  \mathcal{\hat{K}}^{(B)}\,,\ln\left(  \hat{\vec{q}}%
_{1}^{\,\,2}\hat{\vec{q}}_{2}^{\,\,2}\right)  \mathcal{\hat{K}}^{(B)}\right]
|\vec{r}_{1}^{\;\prime},\vec{r}_{2}^{\;\prime}\rangle=\int\frac{d\vec{q}_{1}%
}{2\pi}\frac{d\vec{q}_{2}}{2\pi}\frac{d\vec{k}_{1}}{2\pi}\frac{d\vec{k}_{2}%
}{2\pi}e^{i[\vec{q}_{1}\,\vec{r}_{11^{\prime}}+\vec{q}_{2}\,\vec
{r}_{22^{\prime}}+(\vec{k}\,_{1}+\vec{k}_{2})\,\vec{r}_{1^{\prime}2^{\prime}%
}]}\;
\]%
\[
\ln\left(  \frac{(\vec{q}_{1}-\vec{k}_{1})^{\,\,2}(\vec{q}_{2}+\vec{k}%
_{1})^{\,\,2}}{\vec{q}_{1}^{\,\,2}\vec{q}_{2}^{\,\,2}}\right)  \left[
\frac{1}{\vec{k}_{2}^{\;2}}\left(  \frac{1}{\vec{k}_{1}^{\;2}}+\frac{\vec
{k}_{1}}{\vec{k}_{1}^{\;2}}\left(  \frac{\vec{q}_{2}}{\vec{q}_{2}^{\;2}}%
-\frac{\vec{q}_{1}}{\vec{q}_{1}^{\;2}}\right)  -\frac{\left(  \vec{q}%
_{1}\,\vec{q}_{2}\right)  }{\vec{q}_{1}^{\;2}\vec{q}_{2}^{\;2}}\right)
\right.
\]%
\[
+\frac{1}{\vec{k}_{1}^{\;2}}\left(  \frac{\vec{k}_{2}}{\vec{k}_{2}^{\;2}%
}\left(  \frac{\vec{q}_{2}+\vec{k}_{1}}{(\vec{q}_{2}+\vec{k}_{1})^{\,\,2}%
}-\frac{\vec{q}_{1}-\vec{k}_{1}}{(q_{1}-\vec{k}_{1})^{\,\,2}}\right)
-\frac{(\vec{q}_{1}-\vec{k}_{1})(\vec{q}_{2}+\vec{k}_{1})}{(\vec{q}_{1}%
-\vec{k}_{1})^{\,\,2}(\vec{q}_{2}+\vec{k}_{1})^{\,\,2}}\right)
\]%
\[
+\left(  \frac{\vec{k}_{1}}{\vec{k}_{1}^{\;2}}\left(  \frac{\vec{q}_{2}}%
{\vec{q}_{2}^{\;2}}-\frac{\vec{q}_{1}}{\vec{q}_{1}^{\;2}}\right)  \right)
\left(  \frac{\vec{k}_{2}}{\vec{k}_{2}^{\;2}}\left(  \frac{\vec{q}_{2}+\vec
{k}_{1}}{(\vec{q}_{2}+\vec{k}_{1})^{\,\,2}}-\frac{\vec{q}_{1}-\vec{k}_{1}%
}{(\vec{q}_{1}-\vec{k}_{1})^{\,\,2}}\right)  -\frac{(\vec{q}_{1}-\vec{k}%
_{1})(\vec{q}_{2}+\vec{k}_{1})}{(\vec{q}_{1}-\vec{k}_{1})^{\,\,2}(\vec{q}%
_{2}+\vec{k}_{1})^{\,\,2}}\right)
\]%
\[
-\left.  \frac{\left(  \vec{q}_{1}\,\vec{q}_{2}\right)  }{\vec{q}_{1}%
^{\;2}\vec{q}_{2}^{\;2}}\left(  \frac{\vec{k}_{2}}{\vec{k}_{2}^{\;2}}\left(
\frac{\vec{q}_{2}+\vec{k}_{1}}{(\vec{q}_{2}+\vec{k}_{1})^{\,\,2}}-\frac
{\vec{q}_{1}-\vec{k}_{1}}{(\vec{q}_{1}-\vec{k}_{1})^{\,\,2}}\right)
-\frac{(\vec{q}_{1}-\vec{k}_{1})(\vec{q}_{2}+\vec{k}_{1})}{(\vec{q}_{1}%
-\vec{k}_{1})^{\,\,2}(\vec{q}_{2}+\vec{k}_{1})^{\,\,2}}\right)  \right]
\]%
\[
-\frac{1}{4}\int\frac{d\vec{q}_{1}}{2\pi}\frac{d\vec{q}_{2}}{2\pi}\frac
{d\vec{k}_{1}}{2\pi}\frac{d\vec{k}_{2}}{2\pi}e^{i[\vec{q}_{1}\,\vec
{r}_{11^{\prime}}+\vec{q}_{2}\,\vec{r}_{22^{\prime}}+\vec{k}\,_{1}\,\vec
{r}_{1^{\prime}2^{\prime}}]}\ln\left(  \frac{(\vec{q}_{1}-\vec{k}_{1}%
)^{\,\,2}(\vec{q}_{2}+\vec{k}_{1})^{\,\,2}}{\vec{q}_{1}^{\,\,2}\vec{q}%
_{2}^{\,\,2}}\right)
\]%
\[
\times\left(  \frac{2}{\vec{k}_{2}^{\;2}}+\frac{2\vec{k}_{2}}{\vec{k}%
_{2}^{\;2}}\left(  \frac{\vec{q}_{2}+\vec{k}_{1}-\vec{k}_{2}}{(\vec{q}%
_{2}+\vec{k}_{1}-\vec{k}_{2})^{\,\,2}}+\frac{\vec{q}_{1}-\vec{k}_{1}-\vec
{k}_{2}}{(\vec{q}_{1}-\vec{k}_{1}-\vec{k}_{2})^{\,\,2}}\right)  +\frac
{1}{(\vec{q}_{1}-\vec{k}_{1}-\vec{k}_{2})^{\,\,2}}+\frac{1}{(\vec{q}_{1}%
-\vec{k}_{1}-\vec{k}_{2})^{\,\,2}}\right)
\]%
\begin{equation}
\times\left(  \frac{1}{\vec{k}_{1}^{\;2}}+\frac{\vec{k}_{1}}{\vec{k}_{1}%
^{\;2}}\left(  \frac{\vec{q}_{2}}{\vec{q}_{2}^{\;2}}-\frac{\vec{q}_{1}}%
{\vec{q}_{1}^{\;2}}\right)  -\frac{\left(  \vec{q}_{1}\,\vec{q}_{2}\right)
}{\vec{q}_{1}^{\;2}\vec{q}_{2}^{\;2}}\right)  .
\end{equation}
This expression can be straightforwardly Fourier transformed with the help of
the integrals presented in Ref.~\cite{Fadin:2007de}. Finally we obtain%
\[
4\left(  \frac{\pi^{\,\,2}}{\alpha_{s}N_{c}}\right)  ^{\,2}\langle\vec{r}%
_{1},\vec{r}_{2}|\left[  \mathcal{\hat{K}}^{(B)},\ln\left(  \hat{\vec{q}}%
_{1}^{\,\,2}\hat{\vec{q}}_{2}^{\,\,2}\right)  \mathcal{\hat{K}}^{(B)}\right]
|\vec{r}_{1}^{\;\prime},\vec{r}_{2}^{\;\prime}\rangle
\]%
\[
=4\left(  \frac
{\pi^{\,\,2}}{\alpha_{s}N_{c}}\right)  ^{2}\langle\vec{r}_{1},\vec{r}%
_{2}|\left[  \mathcal{\hat{K}}^{(B)},\ln\left(  \hat{\vec{q}}_{1}^{\,\,2}%
\hat{\vec{q}}_{2}^{\,\,2}\right)  \mathcal{\hat{K}}^{(B)}\right]  _{M}|\vec
{r}_{1}^{\;\prime},\vec{r}_{2}^{\;\prime}\rangle
\]%
\begin{equation}
+\left(  -\frac{2\left(  \vec{r}_{11^{\prime}}\vec{r}_{12^{\prime}}\right)
\ln\left(  \frac{\vec{r}_{11^{\prime}}^{\,\,2}\vec{r}_{12^{\prime}}^{\,\,2}%
}{\vec{r}_{1^{\prime}2^{\prime}}^{\,\,4}}\right)  }{\vec{r}_{11^{\prime}%
}^{\,\,2}\vec{r}_{12^{\prime}}^{\,\,2}\vec{r}_{1^{\prime}2^{\prime}}^{\,\,2}%
}+\delta\left(  \vec{r}_{1^{\prime}2^{\prime}}\right)  \left(  \dots\right)
+\left(  1\leftrightarrow2,1^{\prime}\leftrightarrow2^{\prime}\right)
\right)  ,
\end{equation}
where%
\[
4\left(  \frac{\pi^{\,\,2}}{\alpha_{s}N_{c}}\right)  ^{\,\,2}\langle\vec
{r}_{1},\vec{r}_{2}|\left[  \mathcal{\hat{K}}^{(B)}\,,\ln\left(  \hat{\vec{q}%
}_{1}^{\,\,2}\hat{\vec{q}}_{2}^{\,\,2}\right)  \mathcal{\hat{K}}^{(B)}\right]
_{M}|\vec{r}_{1}^{\;\prime},\vec{r}_{2}^{\;\prime}\rangle=2\pi\delta\left(
\vec{r}_{11^{\prime}}\right)
\]%
\[
\times\left\{  2I\left(  \vec{r}_{12}^{\,\,2},\vec{r}_{22^{\prime}}%
^{\,\,2},\vec{r}_{12^{\prime}}^{\,\,2}\right)  \left(  \frac{\left(  \vec
{r}_{22^{\prime}}\vec{r}_{12^{\prime}}\right)  ^{2}}{\vec{r}_{22^{\prime}%
}^{\,\,2}\vec{r}_{12^{\prime}}^{\,\,2}}-1\right)  -\frac{\left(  \vec
{r}_{22^{\prime}}\vec{r}_{12^{\prime}}\right)  }{2\vec{r}_{22^{\prime}%
}^{\,\,2}\vec{r}_{12^{\prime}}^{\,\,2}}\ln^{2}\frac{\vec{r}_{12}^{\,\,2}}%
{\vec{r}_{12^{\prime}}^{\,\,2}}+\left(  \frac{\left(  \vec{r}_{22^{\prime}%
}\vec{r}_{12^{\prime}}\right)  }{\vec{r}_{22^{\prime}}^{\,\,2}\vec
{r}_{12^{\prime}}^{\,\,2}}-\frac{1}{2\vec{r}_{12^{\prime}}^{\,\,2}}\right)
\ln\frac{\vec{r}_{12}^{\,\,2}}{\vec{r}_{12^{\prime}}^{\,\,2}}\ln\frac{\vec
{r}_{12}^{\,\,2}}{\vec{r}_{22^{\prime}}^{\,\,2}}\right\}
\]%
\[
-2\int d\vec{\rho}\left(  \frac{\vec{r}_{1^{\prime}2^{\prime}}^{\,\,2}}%
{\vec{r}_{1^{\prime}\rho}^{\,\,2}\vec{r}_{2^{\prime}\rho}^{\,\,2}}%
\delta\left(  \vec{r}_{22^{\prime}}\right)  \frac{\left(  \vec{r}_{12}\vec
{r}_{1^{\prime}2}\right)  }{\vec{r}_{11^{\prime}}^{\,\,2}\vec{r}_{1^{\prime}%
2}^{\,\,2}}\ln\frac{\vec{r}_{12}^{\,\,2}}{\vec{r}_{1^{\prime}2}^{\,\,2}}%
-\frac{\delta\left(  \vec{r}_{2\rho}\right)  }{\vec{r}_{2^{\prime}\rho
}^{\,\,2}}\frac{\left(  \vec{r}_{1\rho}\vec{r}_{1^{\prime}\rho}\right)  }%
{\vec{r}_{11^{\prime}}^{\,\,2}\vec{r}_{1^{\prime}\rho}^{\,\,2}}\ln\frac
{\vec{r}_{1\rho}^{\,\,2}}{\vec{r}_{1^{\prime}\rho}^{\,\,2}}-\frac
{\delta\left(  \vec{r}_{22^{\prime}}\right)  }{\vec{r}_{1^{\prime}\rho
}^{\,\,2}}\frac{\left(  \vec{r}_{12}\vec{r}_{\rho2}\right)  }{\vec{r}_{1\rho
}^{\,\,2}\vec{r}_{\rho2}^{\,\,2}}\ln\frac{\vec{r}_{12}^{\,\,2}}{\vec{r}%
_{\rho2}^{\,\,2}}\right)
\]%
\[
+\vec{r}_{12}^{\,\,2}\left(  \frac{\ln\left(  \frac{\vec{r}_{12}^{\,\,4}}%
{\vec{r}_{12^{\prime}}^{\,\,2}\vec{r}_{21^{\prime}}^{\,\,2}}\right)  }%
{2\vec{r}_{11^{\prime}}^{\,\,2}\vec{r}_{22^{\prime}}^{\,\,2}\vec{r}%
_{1^{\prime}2^{\prime}}^{\,\,2}}+\frac{\ln\left(  \frac{\vec{r}_{1^{\prime
}2^{\prime}}^{\,\,4}}{\vec{r}_{11^{\prime}}^{\,\,2}\vec{r}_{22^{\prime}%
}^{\,\,2}}\right)  }{\vec{r}_{11^{\prime}}^{\,\,2}\vec{r}_{12^{\prime}%
}^{\,\,2}\vec{r}_{22^{\prime}}^{\,\,2}}+\frac{\ln\left(  \frac{\vec
{r}_{11^{\prime}}^{\,\,2}\vec{r}_{21^{\prime}}^{\,\,2}\vec{r}_{22^{\prime}%
}^{\,\,2}}{\vec{r}_{12}^{\,\,2}\vec{r}_{1^{\prime}2^{\prime}}^{\,\,4}}\right)
}{\vec{r}_{11^{\prime}}^{\,\,2}\vec{r}_{21^{\prime}}^{\,\,2}\vec{r}%
_{1^{\prime}2^{\prime}}^{\,\,2}}\right)  +\frac{\ln\left(  \frac{\vec
{r}_{11^{\prime}}^{\,\,2}\vec{r}_{21^{\prime}}^{\,\,2}\vec{r}_{22^{\prime}%
}^{\,\,2}}{\vec{r}_{12^{\prime}}^{\,\,2}\vec{r}_{1^{\prime}2^{\prime}}%
^{\,\,4}}\right)  }{\vec{r}_{11^{\prime}}^{\,\,2}\vec{r}_{1^{\prime}2^{\prime
}}^{\,\,2}}\left(  \frac{\vec{r}_{12^{\prime}}^{\,\,2}}{\vec{r}_{22^{\prime}%
}^{\,\,2}}-1\right)
\]%
\begin{equation}
+\frac{\ln\left(  \frac{\vec{r}_{12^{\prime}}^{\,\,2}\vec{r}_{21^{\prime}%
}^{\,\,2}}{\vec{r}_{12}^{\,\,4}}\right)  }{2\vec{r}_{11^{\prime}}^{\,\,2}%
\vec{r}_{22^{\prime}}^{\,\,2}}+\frac{\ln\left(  \frac{\vec{r}_{12}^{\,\,2}%
}{\vec{r}_{12^{\prime}}^{\,\,2}}\right)  }{\vec{r}_{11^{\prime}}^{\,\,2}%
\vec{r}_{12^{\prime}}^{\,\,2}}+\left(  1\leftrightarrow2,1^{\prime
}\leftrightarrow2^{\prime}\right)  . \label{[K_lnq1q2_K]}%
\end{equation}
One can check that this expression vanishes as $\vec{r}_{2}\rightarrow\vec
{r}_{1}$ and hence has the M\"{o}bius property. Next, one needs the integrals
from appendix A to calculate its forward form and to see that it exactly
coincides with the result (\ref{commutator_forward_coord}). Unfortunately it
is clear that this commutator does not eliminate the discrepancy between the kernels.

Another generalization of $\left[  \mathcal{\hat{K}}^{(B)}\,,\ln\left(
\hat{\vec{q}}^{\,\,2}\right)  \mathcal{\hat{K}}^{(B)}\right]  $ to the
nonforward case is $\left[  \mathcal{\hat{K}}^{(B)}\,,\mathcal{\hat{K}}%
^{(B)}\ln\left(  \hat{\vec{q}}_{1}^{\,\,2}\hat{\vec{q}}_{2}^{\,\,2}\right)
\right]  .$ We will calculate its matrix element in the coordinate space via
the identity
\[
\langle\vec{r}_{1},\vec{r}_{2}|\left[  \mathcal{\hat{K}}^{(B)},\,\mathcal{\hat
{K}}^{(B)}\ln\left(  \hat{\vec{q}}_{1}^{\,\,2}\hat{\vec{q}}_{2}^{\,\,2}\right)
\right]  |\vec{r}_{1}^{\,\prime},\vec{r}_{2}^{\,\prime\;}\rangle
\]%
\begin{equation}
=\int
d\vec{\rho}_{1}d\vec{\rho}_{2}\langle\vec{r}_{1},\vec{r}_{2}|\mathcal{\hat{K}%
}^{(B)}|\vec{\rho}_{1},\vec{\rho}_{2}\rangle\langle\vec{\rho}_{1},\vec{\rho
}_{2}|\left[  \mathcal{\hat{K}}^{(B)},\,\ln\left(  \hat{\vec{q}}_{1}%
^{\,\,2}\hat{\vec{q}}_{2}^{\,\,2}\right)  \right]  |\vec{r}_{1}^{\,\prime
},\vec{r}_{2}^{\,\prime\;}\rangle.
\end{equation}
Taking Fourier transform we get%
\begin{equation}
\langle\vec{r}_{1},\vec{r}_{2}|\ln\left(  \hat{\vec{q}}_{1}^{\,\,2}\hat
{\vec{q}}_{2}^{\,\,2}\right)  |\vec{r}_{1}^{\,\prime},\vec{r}_{2}^{\,\prime
\;}\rangle=-\frac{2}{2\pi}\left(  \frac{\delta\left(  \vec{r}_{11^{\prime}%
}\right)  }{\vec{r}_{22^{\prime}}^{\,\,2}}+\frac{\delta\left(  \vec
{r}_{22^{\prime}}\right)  }{\vec{r}_{11^{\prime}}^{\,\,2}}\right)  .
\end{equation}
However, this operator is not M\"{o}bius since its matrix element does not
vanish as $\vec{r}_{1}\rightarrow\vec{r}_{2}$. Yet we can change the matrix
element adding some terms independent of $\vec{r}_{1}$ or of $\vec{r}_{2}$ so
that it satisfies the M\"{o}bius property. We have
\begin{equation}
\langle\vec{r}_{1},\vec{r}_{2}|\ln\left(  \hat{\vec{q}}_{1}^{\,\,2}\hat
{\vec{q}}_{2}^{\,\,2}\right)  _{M}|\vec{r}_{1}^{\,\prime},\vec{r}%
_{2}^{\,\prime\;}\rangle=-\frac{2}{2\pi}\left(  \frac{\delta\left(  \vec
{r}_{11^{\prime}}\right)  }{\vec{r}_{22^{\prime}}^{\,\,2}}-\frac{\delta\left(
\vec{r}_{11^{\prime}}\right)  }{\vec{r}_{12^{\prime}}^{\,\,2}}\right)
+(1\leftrightarrow2,1^{\prime}\leftrightarrow2^{\prime}).
\end{equation}
Thus we arrive to%
\[
\frac{2\pi^{3}}{\alpha_{s}N_{c}}\langle\vec{r}_{1},\vec{r}_{2}|\left[
\mathcal{\hat{K}}^{(B)},\,\ln\left(  \hat{\vec{q}}_{1}^{\,\,2}\hat{\vec{q}%
}_{2}^{\,\,2}\right)  _{M}\right]  |\vec{r}_{1}^{\,\prime},\vec{r}%
_{2}^{\,\prime\;}\rangle=2\pi\delta\left(  \vec{r}_{11^{\prime}}\right)
\frac{(\vec{r}_{12},\vec{r}_{12^{\prime}})}{\vec{r}_{12^{\prime}}^{\,\,2}%
\vec{r}_{22^{\prime}}^{\,\,2}}\ln\left(  \frac{\vec{r}_{12}^{\,\,2}}{\vec
{r}_{12^{\prime}}^{\,\,2}}\right)
\]%
\begin{equation}
+\frac{\vec{r}_{12}^{\,\,2}}{\vec{r}_{11^{\prime}}^{\,\,2}\vec{r}_{21^{\prime
}}^{\,\,2}}\left(  \frac{1}{\vec{r}_{1^{\prime}2^{\prime}}^{\,\,2}}-\frac
{1}{\vec{r}_{22^{\prime}}^{\,\,2}}\right)  +\frac{1}{\vec{r}_{11^{\prime}%
}^{\,\,2}\vec{r}_{1^{\prime}2^{\prime}}^{\,\,2}}\left(  \frac{\vec
{r}_{12^{\prime}}^{\,\,2}}{\vec{r}_{22^{\prime}}^{\,\,2}}-1\right)
+(1\leftrightarrow2,1^{\prime}\leftrightarrow2^{\prime}).
\end{equation}
This matrix element tends to zero as $\vec{r}_{1}\rightarrow\vec{r}_{2}$ and
hence can be convolved with the kernel. Finally we get%
\[
\left(  \frac{2\pi^{2}}{\alpha_{s}N_{c}}\right)  ^{2}\langle\vec{r}_{1}%
,\vec{r}_{2}|\left[  \mathcal{\hat{K}}^{(B)},\,\mathcal{\hat{K}}^{(B)}%
\ln\left(  \hat{\vec{q}}_{1}^{\,\,2}\hat{\vec{q}}_{2}^{\,\,2}\right)
_{M}\right]  |\vec{r}_{1}^{\,\prime},\vec{r}_{2}^{\,\prime\;}\rangle
\]%
\[
=2\int d\vec{\rho}\frac{\vec{r}_{12}^{\,\,2}}{\vec{r}_{1^{\prime}2^{\prime}%
}^{\,\,2}\vec{r}_{1\rho}^{\,\,2}\vec{r}_{2\rho}^{\,\,2}}\left(  \left(
\delta\left(  \vec{r}_{11^{\prime}}\right)  -\delta\left(  \vec{r}_{1^{\prime
}\rho}\right)  \right)  \frac{(\vec{r}_{21^{\prime}}\,\vec{r}_{1^{\prime
}2^{\prime}})}{\vec{r}_{22^{\prime}}^{\,\,2}}\ln\left(  \frac{\vec
{r}_{21^{\prime}}^{\,\,2}}{\vec{r}_{1^{\prime}2^{\prime}}^{\,\,2}}\right)
+\delta\left(  \vec{r}_{11^{\prime}}\right)  \frac{(\vec{r}_{1^{\prime
}2^{\prime}}\,\vec{r}_{1^{\prime}\rho})}{\vec{r}_{2^{\prime}\rho}^{\,\,2}}%
\ln\left(  \frac{\vec{r}_{1^{\prime}\rho}^{\,\,2}}{\vec{r}_{1^{\prime
}2^{\prime}}^{\,\,2}}\right)  \right)
\]%
\begin{equation}
+\vec{r}_{12}^{\,\,2}\left(  \frac{\ln\left(  \frac{\vec{r}_{12^{\prime}%
}^{\,\,2}\vec{r}_{21^{\prime}}^{\,\,2}}{\vec{r}_{12}^{\,\,2}\vec{r}%
_{1^{\prime}2^{\prime}}^{\,\,2}}\right)  }{\vec{r}_{11^{\prime}}^{\,\,2}%
\vec{r}_{22^{\prime}}^{\,\,2}\vec{r}_{1^{\prime}2^{\prime}}^{\,\,2}}+\frac
{\ln\left(  \frac{\vec{r}_{11^{\prime}}^{\,\,2}\vec{r}_{22^{\prime}}^{\,\,2}%
}{\vec{r}_{12}^{\,\,2}\vec{r}_{1^{\prime}2^{\prime}}^{\,\,2}}\right)  }%
{\vec{r}_{11^{\prime}}^{\,\,2}\vec{r}_{21^{\prime}}^{\,\,2}}\left(  \frac
{1}{\vec{r}_{1^{\prime}2^{\prime}}^{\,\,2}}-\frac{1}{\vec{r}_{22^{\prime}%
}^{\,\,2}}\right)  \right)  +\frac{\ln\left(  \frac{\vec{r}_{11^{\prime}%
}^{\,\,2}\vec{r}_{22^{\prime}}^{\,\,2}}{\vec{r}_{12^{\prime}}^{\,\,2}\vec
{r}_{21^{\prime}}^{\,\,2}}\right)  }{\vec{r}_{11^{\prime}}^{\,\,2}\vec
{r}_{1^{\prime}2^{\prime}}^{\,\,2}}\left(  \frac{\vec{r}_{12^{\prime}}%
^{\,\,2}}{\vec{r}_{22^{\prime}}^{\,\,2}}-1\right)  +(1\leftrightarrow
2,1^{\prime}\leftrightarrow2^{\prime}). \label{comm_r}%
\end{equation}
This matrix element has the Mobius property since it vanishes at $\vec{r}%
_{1}=\vec{r}_{2}$.

For the forward case we have
\[
\left(  \frac{2\pi^{2}}{\alpha_{s}N_{c}}\right)  ^{2}\int d\vec{r}_{1^{\prime
}}d\vec{r}_{2^{\prime}}\langle\vec{r}_{1},\vec{r}_{2}|\left[  \mathcal{\hat
{K}}^{(B)},\,\mathcal{\hat{K}}^{(B)}\ln\left(  \hat{\vec{q}}_{1}^{\,\,2}%
\hat{\vec{q}}_{2}^{\,\,2}\right)  _{M}\right]  |\vec{r}_{1}^{\,\prime},\vec
{r}_{2}^{\,\prime\;}\rangle\delta(\vec{r}_{1^{\prime}2^{\prime}}-\vec
{r}^{\,\,\prime})
\]%
\begin{equation}
=-4\pi\frac{\,\vec{r}^{\,\,2}}{(\vec{r}-\vec{r}^{\prime})^{2}\vec
{r}^{\,\,\prime2}}\ln\left(  \frac{\vec{r}^{\,\,2}}{\vec{r}^{\,\,\prime2}%
}\right)  \ln\left(  \frac{\vec{r}^{\,\,2}\vec{r}^{\,\,\prime2}}{(\vec{r}%
-\vec{r}^{\prime})^{4}}\right)  .
\end{equation}
Here $\vec{r}=\vec{r}_{12}$ and we used the integrals from appendix A to
reproduce Eq.~(\ref{commutator_forward}).

\section*{Appendix C}

The  integrals (\ref{difference in omega}) and (\ref{omega M})
were calculated  performing firstly the angular integration
as well as in Ref.~\cite{Kotikov:2000pm},  with the use of   the expansion
over the  Chebyshev polinomials%
\begin{equation}
\frac{1-t^{2}}{1-2tx+t^{2}}=1+2\sum_{n=1}^{\infty}t^{n}T_{n}\left(  x\right)
,\quad\quad\ln\left(  1-2tx+t^{2}\right)  =-2\sum_{n=1}^{\infty}\frac{t^{n}%
}{n}T_{n}\left(  x\right)  ,\qquad\left\vert t\right\vert <1,
\end{equation}%
and the relations
\[
2T_{n}(x)T_{m}(x)=T_{n+m}(x)+T_{n-m}(x),\;\; T_0(x)=1, \qquad T_{-n}(x)=T_{n}(x),
\]
\begin{equation}
\int_{-\pi}^{\pi}\frac{d\phi}{\pi} e^{in\phi}T_m\left(  \cos\phi\right)  =2\int_{0}^{\pi
}\frac{d\phi}{\pi}  e^{in\phi}T_{m}\left(  \cos\phi\right) =\delta_{nm}\left(1+\delta_{n0}\right)~.
\end{equation}
After this the integral (\ref{difference in omega}) can be written as
$-\partial/(\partial\gamma )J_1(n, \gamma)$,  where
\[
J_1(n, \gamma))=\int_0^1\frac{dt}{1-t}\left[t^{\gamma+\frac{n}{2}-1}
\ln t -
2\left(t^{\gamma+\frac{n}{2}-1}+t^{\gamma-\frac{n}{2}-1}-2\right)\ln(1- t)\right.
\]
\begin{equation}
\left.-2t^{\gamma-1}\sum_{l=1}^{n-1}t^{l-\frac{n}{2}}\frac{\left(1-t^{n-l}\right)}{l}
\right]~
+\left(\gamma\leftrightarrow (1-\gamma)\right).\label{J(n gamma)}
\end{equation}
Here and below we assume that $n=|n|$. The integral (\ref{J(n gamma)}) is taken  using the relations
\[
\int_0^1\frac{dt}{1-t}(t^{a-1}-1) =
\psi(1)-\psi(a), \;\;
\]
\[
2\int_0^1\frac{dt}{1-t}(t^{a-1}-1)\ln(1-t)=
\psi'(1)-\psi'(a)+\left(\psi(1)-\psi(a)\right)^2,
\]
\[
-2\int_0^1\frac{dt}{1-t}t^{\gamma-1}\sum_{l=1}^nt^{l-\frac{n}{2}}
\frac{\left(1-t^{n-l}\right)}{l}
+\left(\gamma\leftrightarrow (1-\gamma)\right)
=-2\sum_{l=1}^{n-1} \sum_{m=1}^{n-l}\frac{1}{l}\left(\frac{1}{\gamma-\frac{n}{2}+(l+m-1)}\right.
\]
\begin{equation}
\left.+
\frac{1}{-\gamma+\frac{n}{2}+(l+m-n)}\right)=
2\sum_{l=1}^{n-1} \sum_{k=l}^{n-1}\left(\frac{1}{\gamma-\frac{n}{2}+k}
\frac{1}{\gamma-\frac{n}{2}+k-l}\right)=\sigma_1^2(\gamma, n)-\sigma_2(\gamma, n)~,
\end{equation}
where
\[
\sigma_1(\gamma, n)=\sum_{m=1}^{n} \frac{1}{\gamma-\frac{n}{2}-1+m}, \;\;
\sigma_2(\gamma, n)=\sum_{m=1}^{n} \frac{1}{(\gamma-\frac{n}{2}-1+m)^2}, \;\;
\]
\begin{equation}
\sigma_1(\gamma, n)=-\sigma_1(1-\gamma, n),\;\;\sigma_2(\gamma, n)=\sigma_2(1-\gamma, n).
\end{equation}
After this one can exclude $\psi(a-\frac{n}{2})$ and $\psi'(a-\frac{n}{2})$
exploiting  the properties
\begin{equation}
\psi(a-\frac{n}{2})=\psi(a+\frac{n}{2})-\sigma_1(a, n), \;\;
\psi'(a-\frac{n}{2})=\psi'(a+\frac{n}{2})+\sigma_2(a, n).  \;\;
\label{relation1 for psi}
\end{equation}%
Finally, using the relation
\begin{equation}
\psi'(a+\frac{n}{2})+\psi'(1-a+\frac{n}{2})=-\sigma_2(a, n)+6\psi'(1) +
\left(\psi(1-a+\frac{n}{2})-\psi(a+\frac{n}{2})+\sigma_1(a, n)\right)^2\;\;
\label{relation2 for psi}
\end{equation}%
one obtains
\begin{equation}
J_1(n, \gamma)=2\psi'(1)-\chi^2(n, \gamma), \;\;-\frac{\partial}{\partial\gamma}J_1(n, \gamma)=
2\chi'(n, \gamma)\chi(n, \gamma),
\end{equation}
that gives Eq.~(\ref{difference in omega}).
Let us add for completeness that Eqs.~(\ref{relation1 for psi}) and (\ref{relation2 for psi})
follow from the properties
\[
\psi(x+1)=\frac{1}{x}+\psi(x),\;\;\psi^{\prime}(x+1)=-\frac{1}{x^{2}}%
+\psi^{\prime}(x),\label{psi-relation2}%
\]
\begin{equation}
\psi'(x)+\psi'(1-x)=6\psi'(1)+(\psi(x)-\psi(1-x))^2.
\end{equation}%
In turn, these properties follow from
the definition  $\psi(x)=\left(\ln\Gamma(x)\right)'$
and the properties
\begin{equation}
\Gamma(1+x)=x\Gamma(x), \;\; \Gamma(x)\Gamma(1-x)=\frac{\pi}{\sin\pi x}.
\end{equation}%
The integral (\ref{omega M}) is calculated in the same way. We present here the results of the integration
of the separate terms.  Most of the necessary integrals can be found in
Refs.~\cite{Kotikov:2000pm} and \cite{Balitsky:2008zz}. The integral
\[
J_2(n, \gamma)=\int\frac{d\vec{r}^{\,\,\prime}}{\pi}\left(  \frac{1}{(\vec{r}-\vec
{r}^{\,\,\prime})^{2}}-\frac{1}{\vec{r}^{\,\,\prime2}}\right)  \ln\left(
\frac{(\vec{r}-\vec{r}^{\,\,\prime})^{2}}{\vec{r}^{\,\,\prime2}}\right)
\left(  2\left(  \frac{\vec{r}^{\,\,\prime2}}{\vec{r}^{\,\,2}}\right)
^{\gamma}e^{in(\phi_{\vec{r}^{\,\,\prime}}-\phi_{\vec{r}})}-1\right)
\]%
\begin{equation}
=\chi^{2}(n,\gamma)-\chi^{\prime}(n,\gamma)-\frac{4\gamma\chi(n,\gamma
)}{\gamma^{2}-\frac{n^{2}}{4}}\label{H_result}%
\end{equation}
is calculated quite analogously to $J_1(n, \gamma)$. The calculation of  the integral
\begin{equation}
\int d\vec{r}^{\;\prime}e^{in(\phi_{\vec{r}^{\;\prime}}-\phi_{\vec{r}}%
)}\left(  \frac{\vec{r}^{\;\prime2}}{\vec{r}^{\;2}}\right)  ^{\gamma-1}%
\frac{f_{1}\left(  \vec{r},\vec{r}^{\,\,\prime}\right)  }{2\pi}=-\Phi\left(
n,\gamma\right)  -\Phi\left(  n,1-\gamma\right)  ,
\end{equation}
where $f_{1}\left(  \vec{r},\vec{r}^{\,\,\prime}\right)$ is defined in
Eq.~(\ref{f1})
and $\Phi\left(n,\gamma\right)$  in Eq.~(\ref{Phi1}), does not meet
difficulties after the decomposition
\[
\frac{1} {\left(  \vec x-\vec y\right)  ^{2}\left(
\vec x+\vec y\right)^{2}}=
\frac{1}{2\left(\vec x^{\;2} +\vec y^{\;2}\right)}
\left(\frac{1}{\left(  \vec x-\vec y\right) ^{2}}+\frac{1}{\left(
\vec x+\vec y\right)^{2}}\right)~.
\]
Finally, the integral
\begin{equation}
\int\frac{d\vec{r}^{\,\,\prime}}{\pi}e^{in(\phi_{\vec
{r}^{\,\,\prime}}-\phi_{\vec{r}})}\left(  \frac{\vec{r}^{\;\prime2}}{\vec
{r}^{\;2}}\right) ^{\gamma-1}f_{2}\left(  \vec{r},\vec{r}^{\,\,\prime }\right) =F(n,\gamma)~,
\end{equation}
where $f_{2}\left(  \vec{r},\vec{r}^{\,\,\prime}\right)$ is defined in
Eq.~(\ref{f2}) and $F\left(n,\gamma\right)$  in Eq.~(\ref{F(n,gamma)}),
can be performed using the relations
\begin{equation}
\int_{0}^{\infty}\frac{y^{\alpha}dy}{y+t^{2}}=-\frac{\pi t^{2\alpha}}%
{\sin\left(  \pi\alpha\right)  },\quad\int_{0}^{\infty}dt\,t^{\alpha}%
\ln\left\vert \frac{1-t}{1+t}\right\vert =\frac{\pi\cos\left(  \frac{\pi
\alpha}{2}\right)  }{\left(  \alpha+1\right)  \sin\left(  \frac{\pi\alpha}%
{2}\right)  }%
\end{equation}
after the angular integration.

\end{document}